\documentclass[12pt,draftcls,onecolumn]{IEEEtran}

\ifCLASSINFOpdf

\else

\fi

\usepackage{amssymb}  
\usepackage{graphics} 
\usepackage{epsfig} 
\usepackage{mathptmx} 
\usepackage{times} 
\usepackage{amsmath} 

\usepackage{floatrow}
\usepackage{multirow}

\usepackage{xcolor}
\DeclareMathAlphabet{\mathcal}{OMS}{cmsy}{b}{n}
\DeclareMathAlphabet{\mathcal}{OMS}{cmsy}{m}{n}
\newtheorem{theorem}{\indent Theorem}

\newtheorem{remark}{\indent Remark}
\newtheorem{problem}{\indent Problem}

\newtheorem{assumption}{\indent Assumption}

\begin{document}

\title{Two-stage Estimation for Quantum Detector Tomography: Error Analysis, Numerical and Experimental Results}
\author{Yuanlong~Wang,
Shota~Yokoyama,
Daoyi~Dong,
Ian~R.~Petersen,
Elanor~H.~Huntington,
Hidehiro~Yonezawa

\thanks{This work was supported in part by the Australian Research Council's Discovery Projects funding scheme under Projects DP190101566 and DP180101805, in part by the Australian Research Council Centre of Excellence for Quantum Computation and Communication Technology under Project CE170100012, and in part by the U.S. Office of Naval Research Global under Grant N62909-19-1-2129. (Corresponding author: D. Dong)}
\thanks{Y. Wang, S. Yokoyama and H. Yonezawa are with the School of Engineering and Information Technology, University of New South Wales, Canberra, ACT 2600, Australia, and also with the Centre for Quantum Computation and Communication Technology, Australian Research Council, Canberra, ACT 2600, Australia. Y. Wang is also with Centre for Quantum Dynamics, Griffith University, Brisbane, Queensland 4111, Australia (e-mail: yuanlong.wang.qc@gmail.com; s.yokoyama@adfa.edu.au; h.yonezawa@adfa.edu.au).}%
\thanks{D. Dong is with the School of Engineering and Information Technology, University of New South Wales, Canberra, ACT 2600, Australia (e-mail: daoyidong@gmail.com).}
\thanks{I. R. Petersen is with the Research School of Electrical, Energy and Materials Engineering, Australian National University, Canberra, ACT 2601, Australia (e-mail: i.r.petersen@gmail.com).}
\thanks{E. H. Huntington is with the Research School of Electrical, Energy and Materials Engineering, Australian National University, Canberra, ACT 2600, Australia, and also with the Centre for Quantum Computation and Communication Technology, Australian Research Council, Canberra, ACT 2600, Australia (e-mail: elanor.huntington@anu.edu.au).}
\thanks{Copyright @ 2021 IEEE.  Personal use of this material is permitted.  Permission from IEEE must be obtained for all other uses, in any current or future media, including reprinting/republishing this material for advertising or promotional purposes, creating new collective works, for resale or redistribution to servers or lists, or reuse of any copyrighted component of this work in other works. }
}
\markboth{Accepted version to IEEE Transactions on Information Theory}%
{Shell \MakeLowercase{\textit{et al.}}: Bare Demo of IEEEtran.cls for IEEE Journals}

\maketitle

\begin{abstract}
Quantum detector tomography is a fundamental technique for calibrating quantum devices and performing quantum engineering tasks. In this paper, a novel quantum detector tomography method is proposed. First, a series of different probe states are used to generate measurement data. Then, using constrained linear regression estimation, a stage-1 estimation of the detector is obtained. Finally, the positive semidefinite requirement is added to guarantee a physical stage-2 estimation. This Two-stage Estimation (TSE) method has computational complexity $O(nd^2M)$, where $n$ is the number of $d$-dimensional detector matrices and $M$ is the number of different probe states. An error upper bound is established, and optimization on the coherent probe states is investigated. We perform simulation and a quantum optical experiment to testify the effectiveness of the TSE method.
\end{abstract}


\IEEEpeerreviewmaketitle

\section{Introduction}

\IEEEPARstart{R}{ecent} decades have witnessed the fast development of quantum science and technology, including quantum computation, quantum communication \cite{Nielsen and Chuang 2000} , quantum sensing \cite{quantum sensing}, etc. Measurement, on a quantum entity or using a quantum object, is the connection between the classical (non-quantum) world and the quantum domain, and plays a fundamental role in investigating and controlling a quantum system \cite{Wiseman 2009,berta2014}. For example, quantum computation can be performed through a series of appropriate measurements in certain schemes \cite{raussendorf 2001}. In quantum communication, measurement is a vital part of quantum key distribution \cite{bb84}. In quantum metrology, adaptive measurement can achieve the Heisenberg limit in phase estimation \cite{higgins 2007}.

Since quantum measurement can also be viewed as a class of quantum resource, its investigation and characterization is fundamentally important \cite{learnability}. Quantum detector tomography is a technique to characterize quantum measurement devices \cite{luis 1999,ariano 2004}, and thus paves the way for other estimation tasks like quantum state tomography \cite{paris 2004}-\cite{qi 2015}, Hamiltonian identification \cite{burgarth 2012}-\cite{my 2016} and quantum process tomography \cite{jf and hradil 2001}-\cite{ji2008}.

The investigation of protocols for quantum detector tomography dates back to \cite{mle 2001}, where the Maximum Likelihood Estimation (MLE) method is employed to reconstruct an unknown POVM detector. As one of the most widely recognized methods \cite{paris 2004,auria 2011}, MLE can preserve the positivity and completeness of the detector, but it is difficult to characterize the error and computational complexity. Phase-insensitive detectors correspond to diagonal matrices in the photon number basis and are thus relatively straightforward to be reconstructed. Ref. \cite{grandi 2017} modelled this problem as a linear-regression problem and obtained a least squares solution. In \cite{lundeen 2009,feito 2009}, phase-insensitive detector tomography was modelled as a convex quadratic optimization problem and an efficient numerical solution was obtained. This method was also experimentally tested in \cite{natarajan 2013,brida 2012}, and then was developed in \cite{lundeen 2012a} and \cite{lundeen 2012b} to model phase-sensitive detector tomography as a recursive constrained convex optimization problem, where the unknown parameters are recursively estimated. For phase-insensitive detectors with a large linear loss, an extension of detector tomography was introduced in \cite{renema 2012} and tested on a superconducting multiphoton nanodetector.

In this paper, we propose a novel quantum detector tomography protocol, which is applicable to both phase-insensitive and general phase-sensitive detectors. We first input a series of different states (probe states) to the detector, and then collect all the measurement data. The forthcoming algorithm mainly consists of two stages: in the first stage, we find a constrained least square estimate, which corresponds to a Hermitian estimate satisfying the completeness constraint. However, this estimate can be non-physical; i.e., the estimated detectors may have negative eigenvalues. Hence, in the second stage we further design a series of matrix transformations preserving the Hermitian and completeness constraint to find a physical approximation based on the result in the first stage, and thus obtain the final physical estimate. Our Two-stage Estimation (TSE) method has computational complexity $O(nd^2M)$, where $n$ and $d$ are the number and dimension of the detector matrices, respectively, and $M$ is the number of different probe states. This theoretical characterization of the computational complexity is not common in other detector tomography methods. We further prove an error upper bound $O(\frac{d^5n^2}{N})$ on the condition that the probe states are optimal (if not optimal, the specific form of the bound is also given in Sec. \ref{secerror}), where $N$ is the total copy number of probe states. We then investigated optimization of the types of coherent probe states and the size of their sampling square. We perform numerical simulation to validate the theoretical analysis and compare our algorithm with MLE method. Finally, we slightly modify our method to cater to a practical experiment situation, and we perform quantum optical experiments using two-mode coherent states to testify the effectiveness of our method.

This paper is organized as follows. In Section \ref{sec2}, we introduce some preliminary knowledge about quantum physics and formulate our estimation problem. In Section \ref{sec3}, we present the procedures of our TSE method and analyze the computational complexity. An upper bound for the estimation error of TSE is established in Section \ref{secerror}. Section \ref{sec5} investigates the optimization of the coherent probe states. Section \ref{secsimu} presents the numerical simulation results to verify the theoretical analysis in Section \ref{secerror} and Section \ref{sec5}, and to compare our method with MLE. Section \ref{secexp} modifies the TSE method according to our practical physical setting and presents the experimental results. Section \ref{secfinal} concludes this paper.

\textit{Notation:}
$A\geq 0$ means $A$ is positive semidefinite. $A^\dagger$ is the conjugation ($*$) and transpose ($T$) of $A$. $I$ is the identity matrix. $\mathbb{R}$ and $\mathbb{C}$ are the real and complex domains, respectively. $\otimes$ is the tensor product. $\oplus$ is the matrix direct sum. $\text{vec}$ is the column vectorization function. $||\cdot||$ is the Frobenius norm. $\delta$ is the Kronecker delta function. $\text{i}=\sqrt{-1}$. $\text{diag}(X)$ has two effects: it outputs a diagonal matrix with the diagonal elements being the elements in $X$ if $X$ is a vector, or sets all the non-diagonal elements in $X$ to be zero if $X$ is a square matrix. $\hat X$ denotes the estimation of variable $X$. For any positive semidefinite $X_{d\times d}$ with spectral decomposition $X=UPU^\dagger$, define $\sqrt{X}$ or $X^{\frac{1}{2}}$ as $U\text{diag}(\sqrt{P_{11}},\sqrt{P_{22}},...,\sqrt{P_{dd}})U^\dagger$.

\section{Preliminaries and Problem formulation}\label{sec2}
\subsection{Quantum state and measurement}
For a $d$-dimensional quantum system, its state is usually described by a $d\times d$ Hermitian matrix $\rho$, which should be positive semidefinite and satisfy $\text{Tr}(\rho)=1$. When $\rho$ is a pure state (satisfying $\text{Tr}(\rho^2)=1$), we have $\rho=|\psi\rangle\langle\psi|$ where $|\psi\rangle$ is a complex vector on the $d$-dimensional underlying Hilbert space. In this case, we usually identify $\rho$ with $|\psi\rangle$. Otherwise, $\rho$ is called a mixed state, and can be expanded using pure states $\{|\psi_i\rangle\}$: $\rho=\sum_i c_i|\psi_i\rangle\langle\psi_i|$ where $c_i\in\mathbb{R}$ and $\sum_i c_i=1$. The evolution of a pure state $|\psi\rangle$ is described by the Schr\"{o}dinger equation
$$\text{i}\hbar\frac{d}{d t}|\psi(t)\rangle=\mathcal{H}|\psi(t)\rangle,$$
where $\hbar$ is the reduced Planck constant and $\mathcal{H}$ is the system Hamiltonian.

One of the most common quantum measurement methods is the positive operator valued measure (POVM), and quantum detectors are devices to realize a POVM, especially in the optical domain. A set of POVM elements is a set of operators $\{P_i\}$ satisfying the \textit{completeness constraint} $\sum_i P_i=I$ and each $P_i$ is Hermitian and positive semidefinite. In the case when each operator $P_i$ is infinite dimensional, they are usually truncated at a finite dimension $d$ in practice. When the measurements corresponding to operators $\{P_i\}$ are performed on $\rho$, the probability of obtaining the $i$-th result is given by the Born Rule $$p_i=\text{Tr}(P_i\rho).$$
From the completeness constraint, we thus have $\sum_i p_i=1$. In practical experiments, suppose that $N$ (also called the resource number) identical copies of $\rho$ are prepared and the $i$-th results occur $N_i$ times. Then $N_i/N$ is the experimental estimation of the true value $p_i$. The measurement apparatus is the physical realization of a quantum detector, and $\{P_i\}$ is the mathematical representation. We thus directly call $\{P_i\}$ a quantum detector in this paper.

\subsection{Problem formulation}
The technique to deduce an unknown detector from known quantum states and measurement results is called quantum detector tomography. Suppose the true values of a set for a detector are $\{P_i\}_{i=1}^{n}$ such that $\sum_{i=1}^nP_i=I$ with each $P_i$ Hermitian and positive semidefinite. We design a series of different quantum states $\rho_j$ (called \textit{probe states}) and record the measurement results $\hat p_{ij}$ as the estimate of $p_{ij}=\text{Tr}(P_i\rho_j)$. Assume that $M$ different types of probe states are employed and their total number of copies is $N$. Also assume different probe states use the same number of copies, which is $N/M$. We then aim to solve the following optimization problem:
\begin{problem}\label{prob0}
Given experimental data $\{\hat p_{ij}\}$, solve $\min_{\{\hat P_i\}}\sum_{i=1}^{n}\sum_{j=1}^{M} [\hat p_{ij}-\text{Tr}(\hat P_i\rho_j)]^2$ such that $\sum_{i=1}^n \hat P_i=I$ and $\hat P_i\geq0$ for $1\leq i\leq n$.
\end{problem}

\section{Estimation Algorithm}\label{sec3}

\subsection{Stage-1 approximation--constrained LRE}\label{seclre}
We first parameterize the detector and the input (probe) states. Let $\{\Omega_i\}_{i=1}^{d^2}$ be a complete set of $d$-dimensional traceless Hermitian matrices except $\Omega_1=I/\sqrt{d}$, and they satisfy $\text{Tr}(\Omega_i^\dagger\Omega_j)=\delta_{ij}$, where $\delta_{ij}$ is the Kronecker function. Denote the detector by $\{P_i\}$ which are positive semidefinite and $\sum_{i=1}^n P_i=I$. Let $\rho_j$ be a series of input probe states. Then we can parameterize the detector and probe states as
\begin{equation}\label{eqa1}
P_i=\sum_{a=1}^{d^2}\theta_a^{(i)}\Omega_a,
\end{equation}
\begin{equation}\label{eqa2}
\rho_j=\sum_{b=1}^{d^2}\phi_b^{(j)}\Omega_b,
\end{equation}
where $\theta_a^{(i)}\triangleq\text{Tr}(P_i\Omega_a)$ and $\phi_b^{(j)}\triangleq\text{Tr}(\rho_j\Omega_b)$ are real. When $\rho_j$ is inputted, the probability to obtain the result corresponding to $P_i$ is calculated according to Born's rule as
\begin{equation}\label{eqa3}
p_{ij}=\text{Tr}(P_i\rho_j).
\end{equation}
Substituting (\ref{eqa1}) and (\ref{eqa2}) into (\ref{eqa3}), we obtain
$$p_{ij}=\sum_{a=1}^{d^2}\phi_a^{(j)}\theta_a^{(i)}= \Phi_j^T\Theta_i,$$
where $\Phi_j\triangleq(\phi_1^{(j)},\phi_2^{(j)},...,\phi_{d^2}^{(j)})^T$ and $\Theta_i\triangleq(\theta_1^{(i)},\theta_2^{(i)},...,\theta_{d^2}^{(i)})^T$. Suppose when estimating $\hat p_{ij}$, the outcome for $P_i$ appears $n_{ij}$ times, then $\hat p_{ij}=n_{ij}/(N/M)$. Denote the error as $e_{ij}=\hat p_{ij}- p_{ij}$. According to the central limit theorem, $e_{ij}$ converges in distribution to a normal distribution with mean zero and variance $(p_{ij}-p_{ij}^2)/(N/M)$. We thus have the linear regression equation
$$\hat p_{ij}=\Phi_j^T\Theta_i+e_{ij}.$$
Let $\Theta=(\Theta_1^T,\Theta_2^T,...,\Theta_n^T)^T$, which is the vector of all the unknown parameters to be estimated. Collect the parametrization of the probe states as $X_0=(\Phi_1,\Phi_2,...,\Phi_M)^T$. Let $\hat Y=(\hat p_{11},\hat p_{12},...,\hat p_{1M},\hat p_{21},\hat p_{22},...,\hat p_{2M},...,\hat p_{nM})^T$, $X=I_n\otimes X_0$, $e=(e_{11},e_{12},...,e_{1M},e_{21},e_{22},...,e_{2M},...,e_{nM})^T$, $H=(1,1,...,1)_{1\times n}\otimes I_{d^2}$, $D_{d^2\times 1}=(\sqrt{d},0,...,0)^T$. Then the regression equations can be rewritten in a compact form:
\begin{equation}\label{eqa06}
\hat Y=X\Theta+e,
\end{equation}
with a linear constraint
\begin{equation}\label{eqa07}
H\Theta=D.
\end{equation}
Now Problem \ref{prob0} can be transformed into the following equivalent form:
\begin{problem}\label{prob01}
Given experimental data $\hat Y$, solve $\min_{\{\hat P_i\}} ||\hat Y-X\hat\Theta||^2$ such that $H\hat\Theta=D$ and $\hat P_i\geq0$ for $1\leq i\leq n$, where $\hat \Theta$ is the parametrization of ${\{\hat P_i\}}$ via (\ref{eqa1}).
\end{problem}

Problem \ref{prob01} is difficult to solve directly. Hence, we split it into two approximate subproblems:
\addtocounter{problem}{-1}
\renewcommand{\theproblem}{\arabic{problem}$.1$}
\begin{problem}\label{subproblem1}
Given experimental data $\hat Y$, solve $\min_{\{\hat E_i\}} ||\hat Y-X\hat\Theta||^2$ such that $H\hat\Theta=D$, where $\hat \Theta$ is the parametrization of ${\{\hat E_i\}}$ via (\ref{eqa1}).
\end{problem}
\renewcommand{\theproblem}{\arabic{problem}}
\addtocounter{problem}{-1}
\renewcommand{\theproblem}{\arabic{problem}$.2$}
\begin{problem}\label{subproblem2}
Given $\sum_{i=1}^n \hat E_i=I$, solve $\min_{\{\hat P_i\}} \sum_i||\hat E_i-\hat P_i||^2$ such that $\sum_i \hat P_i=I$ and $\hat P_i\geq0$ for $1\leq i\leq n$.
\end{problem}
\renewcommand{\theproblem}{\arabic{problem}}

Problem \ref{subproblem1} is a linear regression problem with a linear constraint, and it can be solved analytically via the Constrained Least Squares (CLS) method \cite{regression}. Assume the input states have enough diversity such that $X^TX$ is nonsingular. This indicates $M\geq d^2$ for general complete probe-state sets. The standard CLS solution is \cite{regression}
\begin{equation}\label{eqa08}
\hat \Theta_{CLS}=\hat \Theta_{LS}-(X^TX)^{-1}H^T[H(X^TX)^{-1}H^T]^{-1}(H\hat \Theta_{LS}-D),
\end{equation}
where $\hat \Theta_{LS}$ is unconstrained least square solution
\begin{equation}\label{eqa09}
\hat \Theta_{LS}=(X^TX)^{-1}X^T\hat Y.
\end{equation}

To further reduce the computational burden, we can simplify the form of (\ref{eqa08}) and (\ref{eqa09}). Let $Z_0=(X_0^TX_0)^{-1}$. Then $(X^TX)^{-1}=I_n\otimes Z_0$, and
$$[H(X^TX)^{-1}H^T]^{-1}=[H(I_n\otimes Z_0)H^T]^{-1}=(nZ_0)^{-1}=\frac{1}{n}Z_0^{-1}.$$
Eq. (\ref{eqa09}) is in fact
$$\hat \Theta_{LS}=(I_n\otimes Z_0)(I_n\otimes X_0^T)\hat Y=(I_n\otimes Z_0 X_0^T)\hat Y.$$
Also (\ref{eqa08}) is
\begin{equation}\label{eqa10}
\begin{array}{rl}
&\ \ \ \ \hat \Theta_{CLS}\\
&=(I_n\otimes Z_0 X_0^T)\hat Y-(I_n\otimes Z_0)\left(\begin{matrix}
I_{d^2}\\
\vdots\\
I_{d^2}\\
\end{matrix}\right)\frac{1}{n}Z_0^{-1}\\
&\ \ \ \ \cdot[(I_{d^2}\cdots I_{d^2})(I_n\otimes Z_0 X_0^T)\hat Y-D]\\
&=(I_n\otimes Z_0 X_0^T)\hat Y-\frac{1}{n}\left(\begin{matrix}
Z_0\\
\vdots\\
Z_0\\
\end{matrix}\right)Z_0^{-1}[(Z_0X_0^T\cdots Z_0X_0^T)\hat Y-D]\\
&=(I_n\otimes Z_0 X_0^T)\hat Y-\frac{1}{n}\left(\begin{matrix}
I_{d^2}\\
\vdots\\
I_{d^2}\\
\end{matrix}\right)[(Z_0X_0^T\cdots Z_0X_0^T)\hat Y-D]\\
&=(I_n\otimes Z_0 X_0^T)\hat Y-\frac{1}{n}\left(\begin{matrix}
Z_0X_0^T&\cdots& Z_0X_0^T\\
\vdots& &\vdots\\
Z_0X_0^T&\cdots &Z_0X_0^T\\
\end{matrix}\right)\hat Y+\frac{1}{n}\left(\begin{matrix}
D\\
\vdots\\
D\\
\end{matrix}\right).\\
\end{array}
\end{equation}
We then partition $\hat Y$ as $\hat Y^T=(\hat Y_1^T,\hat Y_2^T,...,\hat Y_n^T)^T$ where $\hat Y_i=(\hat p_{i1},\hat p_{i2},...,\hat p_{iM})^T$ for $1\leq i\leq n$. Denote $Y_0=((1,...,1)_{1\times M})^T=\sum_i\hat Y_i$. We continue transforming (\ref{eqa10}) as
\begin{equation}\label{eqa13}
\begin{array}{rl}
\hat \Theta_{CLS}&=\left(\begin{matrix}
Z_0 X_0^T & &\\
&\ddots &\\
& &Z_0 X_0^T\\
\end{matrix}\right)\left(\begin{matrix}
\hat Y_1\\
\vdots\\
\hat Y_n\\
\end{matrix}\right)\\
&\ \ \ \ -\frac{1}{n}\left(\begin{matrix}
Z_0X_0^T&\cdots& Z_0X_0^T\\
\vdots& &\vdots\\
Z_0X_0^T&\cdots &Z_0X_0^T\\
\end{matrix}\right)\left(\begin{matrix}
\hat Y_1\\
\vdots\\
\hat Y_n\\
\end{matrix}\right)+\frac{1}{n}\left(\begin{matrix}
D\\
\vdots\\
D\\
\end{matrix}\right)\\
&=\left(\begin{matrix}
Z_0X_0^T\hat Y_1\\
\vdots\\
Z_0X_0^T\hat Y_n\\
\end{matrix}\right)-\frac{1}{n}\left(\begin{matrix}
Z_0X_0^T\sum_i\hat Y_i\\
\vdots\\
Z_0X_0^T\sum_i\hat Y_i\\
\end{matrix}\right)+\frac{1}{n}\left(\begin{matrix}
D\\
\vdots\\
D\\
\end{matrix}\right)\\
&=\left(\begin{matrix}
(X_0^TX_0)^{-1}X_0^T(\hat Y_1-\frac{1}{n}Y_0)+\frac{1}{n}D\\
\vdots\\
(X_0^TX_0)^{-1}X_0^T(\hat Y_n-\frac{1}{n}Y_0)+\frac{1}{n}D\\
\end{matrix}\right).\\
\end{array}
\end{equation}
Compared with (\ref{eqa08}) and (\ref{eqa09}), Eq. (\ref{eqa13}) is a faster way to calculate $\hat \Theta_{CLS}$.

Let $\hat\Theta_{CLS}=(\hat\Theta_1^T,...,\hat\Theta_n^T)^T$ and $\hat\Theta_i^T=(\hat\theta_1^{(i)},...,\hat\theta_{d^2}^{(i)})$. From $\hat \Theta_{CLS}$, we can obtain a stage-1 estimate $\hat E_i=\sum_{a=1}^{d^2}\hat \theta_a^{(i)}\Omega_a$. The error $||\hat E_i-E_i||$ between the estimate $\hat E_i$ and its true value $E_i$ will be referred to as \textit{the CLS error} in the rest of this paper. Note that the positive semidefiniteness requirement on $\hat E_i$ is not considered at this stage, and $\hat E_i$ may have negative eigenvalues. Hence, we need to further adjust $\hat E_i$ to obtain a physical estimate.

\subsection{Difference decomposition}\label{sec202}

Now we begin to solve Problem \ref{subproblem2}. First we decompose each $\hat E_i$ as the difference of two positive semidefinite matrices $\hat F_i$ and $\hat G_i$: $\hat E_i=\hat F_i-\hat G_i$. There are infinitely many such decompositions, because a new decomposition will be obtained once another positive semidefinite matrix is added to $\hat F_i$ and $\hat G_i$. We hope to view $\hat G_i$ as small disturbance, and we thus seek a decomposition method to minimize the norm of $\hat G_i$.

For each $\hat E_i$, we perform a spectral decomposition to obtain $\hat E_i=\hat W_i\hat K_i\hat W_i^\dagger$, where $\hat W_i$ is unitary and $\hat K_i$ is real diagonal. We have $$\hat K_i=\hat W_i^\dagger\hat F_i\hat W_i- \hat W_i^\dagger\hat G_i\hat W_i.$$ Denote the optimal decomposition solution as $\hat F_i^o$ and $\hat G_i^o$. We assert that both $\hat W_i^\dagger\hat F_i^o\hat W_i$ and $\hat W_i^\dagger\hat G_i^o\hat W_i$ must be diagonal. Otherwise, we note that $\text{diag}(\hat W_i^\dagger\hat F_i^o\hat W_i)-\text{diag}(\hat W_i^\dagger\hat G_i^o\hat W_i)$ still equals to $\hat K_i$. Since $\hat W_i^\dagger\hat F_i^o\hat W_i$ is positive semidefinite, all of its diagonal elements are thus nonnegative. This indicates that $\text{diag}(\hat W_i^\dagger\hat F_i^o\hat W_i)$ is also positive semidefinite. Similarly, $\text{diag}(\hat W_i^\dagger\hat G_i^o\hat W_i)$ is also positive semidefinite. Hence, $\text{diag}(\hat W_i^\dagger\hat F_i^o\hat W_i)$ and $\text{diag}(\hat W_i^\dagger\hat G_i^o\hat W_i)$ are also feasible solutions. Since $||\text{diag}(\hat W_i^\dagger\hat G_i^o\hat W_i)||<||\hat W_i^\dagger\hat G_i^o\hat W_i||$, this contradicts the assumption that $\hat G_i^o$ is the optimal solution. Therefore, $\hat W_i^\dagger\hat F_i^o\hat W_i$ and $\hat W_i^\dagger\hat G_i^o\hat W_i$ must be diagonal. We then have $$\min ||\hat G_i||^2=\min ||\hat W_i^\dagger\hat G_i^o\hat W_i||^2=\sum_j\min (\hat W_i^\dagger\hat G_i^o\hat W_i)_{jj}^2,$$ and we can consider its elements: $(\hat K_i)_{jj}=(\hat W_i^\dagger\hat F_i\hat W_i)_{jj}- (\hat W_i^\dagger\hat G_i\hat W_i)_{jj}$. If $(\hat K_i)_{jj}>0$, we should take $(\hat W_i^\dagger\hat F_i\hat W_i)_{jj}=(\hat K_i)_{jj}$ and $(\hat W_i^\dagger\hat G_i\hat W_i)_{jj}=0$; if $(\hat K_i)_{jj}\leq0$, we should take $(\hat W_i^\dagger\hat F_i\hat W_i)_{jj}=0$ and $(\hat W_i^\dagger\hat G_i\hat W_i)_{jj}=-(\hat K_i)_{jj}$.

The optimal decomposition can be obtained through the following procedure. Assume there are $\hat n_i$ nonpositive eigenvalues for $\hat E_i$, and they are in decreasing order in $\text{diag}(\hat K_i)$. Let
\begin{equation}\label{eq2}
\hat F_i=\hat W_i\text{diag}[(\hat K_i)_{11},(\hat K_i)_{22},...,(\hat K_i)_{(d-\hat n_i)(d-\hat n_i)},0,...,0]\hat W_i^\dagger
\end{equation}
and
\begin{equation}
\begin{array}{rl}
\hat G_i&=-\hat W_i\text{diag}[0,...,0,(\hat K_i)_{(d-\hat n_i+1)(d-\hat n_i+1)},\\
&\ \ \ \ (\hat K_i)_{(d-\hat n_i+2)(d-\hat n_i+2)},...,(\hat K_i)_{dd}]\hat W_i^\dagger.\\
\end{array}\nonumber
\end{equation}
Then we know $\hat F_i\geq0$, $\hat G_i\geq0$ and $\hat E_i=\hat F_i-\hat G_i$, and this $\hat G_i$ has the least norm. Also note that for the true values we have $E_i=F_i=P_i$.

\subsection{Stage-2 approximation}\label{sec203}
From $I=\sum_i\hat E_i=\sum_i\hat F_i-\sum_i\hat G_i$, we have
\begin{equation}\label{eq1}
I+\sum_i\hat G_i=\sum_i\hat F_i.
\end{equation}
Since each $\hat G_i$ is positive semidefinite, we can decompose
\begin{equation}\label{eq03}
I+\sum_i\hat G_i=\hat C\hat C^\dagger.
\end{equation}
Then Eq. (\ref{eq1}) is transformed into
$$\sum_i\hat C^{-1}\hat F_i(\hat{C}^{\dagger})^{-1}=I.$$
We let $\hat A_i=\hat C^{-1}\hat F_i(\hat{C}^{\dagger})^{-1}$, and then each $\hat A_i$ is positive semidefinite and their sum is the identity. Hence, $\{\hat A_i\}$ is a genuine estimate of the detector and we call $\{\hat A_i\}$ the stage-2 approximation. A further optimization is needed in order to obtain the final estimation result in the following.

\subsection{Unitary optimization}\label{sec204}

When decomposing $I+\sum_i\hat G_i=\hat C\hat C^\dagger$, there is in fact another degree of freedom. For any unitary $\hat U$, it holds that $\hat C\hat C^\dagger=\hat C\hat U\hat U^\dagger\hat C^\dagger$. Therefore, $\hat U^\dagger \hat A_i\hat U$ can also be an estimate of the detector. There might be multiple ways of defining the cost function to choose an appropriate $\hat{U}$, and here we illustrate one effective way. We hope to choose a $\hat U$ such that the effect of $\hat C$ is (partly) neutralized; i.e., $\hat{U}^\dagger\hat{C}^{-1}(\cdot)(\hat{C}^{\dagger})^{-1}\hat{U}$ has minimal effect on $\hat{F}_i$. This requires $\hat{C}\hat{U}$ to be close to $I$, and hence, we aim to minimize $||\hat C\hat U-I||$.

We have
\begin{equation}
\begin{array}{rl}
||\hat C\hat U-I||^2&=\text{Tr}[(\hat C\hat U-I)(\hat U^\dagger\hat C^\dagger-I)]\\
&=d+\text{Tr}(\hat C\hat C^\dagger)-\text{Tr}(\hat C\hat U+\hat U^\dagger\hat C^\dagger).\\
\end{array}\nonumber
\end{equation}
Let $L=-\text{Tr}(\hat C\hat U+\hat U^\dagger\hat C^\dagger)+\text{Tr}[(\Lambda+\Lambda^\dagger)(\hat U\hat U^\dagger-I)]$ where $\Lambda$ is a Lagrange multiplier matrix and the term $\text{Tr}[(\Lambda+\Lambda^\dagger)(\hat U\hat U^\dagger-I)]$ amounts to the unitary requirement on $\hat{U}$. By partial differentiation we have
$$\frac{\partial L}{\partial \hat U^*}=-\hat C^\dagger+(\Lambda+\Lambda^\dagger)\hat U=0.$$
Therefore,
\begin{equation}\label{eq07}
\hat C^\dagger\hat U^\dagger=\Lambda+\Lambda^\dagger=\hat U\hat C.
\end{equation}
We perform a singular value decomposition to obtain $\hat C=\hat U_\alpha\hat S\hat U_\beta^\dagger$ where $\hat U_\alpha$ and $\hat U_\beta$ are unitary and $\hat S$ is diagonal and positive semidefinite. It is straightforward to verify that
$$\sqrt{\hat C\hat C^\dagger}\hat C^{-1}=\hat U_{\beta}\hat U_{\alpha}^\dagger.$$
Let $\hat U_\gamma=\hat U_\beta^\dagger\hat U\hat U_\alpha$. Then (\ref{eq07}) is now equivalent to
\begin{equation}\label{eq08}
\hat S\hat U_\gamma^\dagger =\hat U_\gamma\hat S.
\end{equation}
Thus we have $$\hat U_\gamma\hat S^2\hat U_\gamma^\dagger =\hat U_\gamma\hat S\hat U_\gamma\hat S=\hat S\hat U_\gamma^\dagger \hat U_\gamma \hat S=\hat S^2.$$ Therefore, $\hat U_\gamma\hat S^2\hat U_\gamma^\dagger$ is the spectral decomposition of $\hat S^2$. Since the probability for $\hat S$ to be degenerate is zero, we know $\hat S^2$ is nondegenerate. Thus we have $\hat U_\gamma=\text{diag}(\text{e}^{\text{i}\theta_1},\text{e}^{\text{i}\theta_2},...,\text{e}^{\text{i}\theta_d})$ where $\theta_j\in [0,2\pi)$ for $1\leq j\leq d$, which indicates that $\hat U_\gamma$ and $\hat S$ commutate. From (\ref{eq08}), we then have $\hat S=\hat U_\gamma\hat S\hat U_\gamma=\hat S\hat U_\gamma^2$. When the resource number $N$ is large enough, $\hat{C}\hat{C}^\dagger$ will be close enough to the identity matrix, and $\hat C$ will be close to a unitary matrix (thus nonsingular) and we can view $\hat S$ as nonsingular. We thus have $\hat U_\gamma^2=I$, which indicates $\hat U_\gamma=\text{diag}(\pm1,\pm1,...,\pm1)$. We further have $L=-2\text{Tr}(\hat U_\gamma\hat S)$, which indicates $\hat U_\gamma=I$. Therefore, the optimal solution is
\begin{equation}\label{eq9}
\hat U=\hat U_\beta\hat U_\alpha^\dagger=\sqrt{\hat C^\dagger\hat C}\hat C^{-1}.
\end{equation}
Hence, the final estimation is $\hat P_i=\hat U^\dagger \hat A_i\hat U$ where $\hat U$ is determined through (\ref{eq9}). The error $||\hat P_i-P_i||$ will be referred to as \textit{the final (estimation) error}, in contrast to the CLS error $||\hat E_i-E_i||$.

\subsection{General procedure and computational complexity}\label{sec205}
We now generalize the procedure of our TSE algorithm and analyze its computational complexity. In this paper, we do not consider the time spent on experiments, since it depends on the experimental realization. In the following, we briefly summarize each step and illustrate their corresponding computational complexity.

\textbf{Step 1. Stage-1 Approximation.} Choose basis sets $\{\Omega_i\}$ and probe states $\rho_j$ and calculate $\Phi_j$. Then perform measurement experiments to collect data $\hat p_{ij}$. Obtain the constrained least square solution from (\ref{eqa13}) and construct the stage-1 approximation $\hat E_i=\sum_a\hat\theta_a^{(i)}\Omega_a$. In (\ref{eqa13}), both $(X_0^TX_0)^{-1}X_0^T$ and $D$ can be calculated offline prior to the experiments, and the remaining online calculation has computational complexity $O(nd^2M)$.

\textbf{Step 2. Difference Decomposition.} Perform spectral decomposition on each $\hat E_i$ and obtain $\hat E_i=\hat F_i-\hat G_i$. Since the computational complexity of spectral decomposition is cubic in the dimension of a Hermitian matrix \cite{schur2}, this step has total computational complexity $O(nd^3)$.

\textbf{Step 3. Stage-2 Approximation.} The transformation of $\sum_i\hat F_i$ into $\hat C\hat C^\dagger$ can be accomplished by spectral decomposition. Then, we obtain the stage-2 approximation $\hat A_i=\hat C^{-1}\hat F_i(\hat{C}^{\dagger})^{-1}$. The complexity of this step is $O(nd^3)$.

\textbf{Step 4. Unitary Optimization.} Calculate the global unitary matrix $\hat U$ according to (\ref{eq9}) and obtain the final estimate $\hat P_i=\hat U^\dagger \hat A_i\hat U$. This step has computational complexity $O(nd^3)$.

Since $M\ge d^2$ for general complete probe-state sets, we have $nd^3\leq nd^2M$. Hence, our algorithm has total computational complexity $O(nd^2M)$.

\section{Error Analysis}\label{secerror}
In this section, we present a theoretical upper bound for the final estimation error of our TSE algorithm. It is necessary to first characterize the probe states.
\begin{assumption}\label{assum1}
The probe states used are \textit{optimal} \cite{qi 2013,my 2016}; i.e., they are $d$-dimensional pure states and $X_0^TX_0=c_0I$ where $c_0\in\mathbb{R}$. From \cite{my 2016}, we have the following characterization:
\begin{equation}\label{assum2}
\frac{M}{4N}\text{Tr}[(X_0^TX_0)^{-1}]= O(\frac{d^4}{N}).
\end{equation}
\end{assumption}

Let $\text{E}(\cdot)$ denote the expectation w.r.t. all possible measurement results. We present the following theorem to characterize the estimation error:
\begin{theorem}\label{the1}
Under Assumption \ref{assum1}, the final estimation error of our algorithm $\text{E}(\sum_i||\hat P_i-P_i||^2)$ scales as $O(\frac{d^5n^2}{{N}})$, where $d$ is the system dimension, $n$ is the number of detector POVM elements and $N$ is the total number of resources.
\end{theorem}

\begin{IEEEproof}
We prove the conclusion through analyzing the error in each step of our algorithm.
\subsection{Error in stage-1 approximation}

For simplicity, let $Z=(X^TX)^{-1}=I\otimes(X_0^TX_0)^{-1}$. The estimation error for constrained LRE is
\begin{equation}
\begin{array}{rl}
&\ \ \ \ \text{E}[||\hat \Theta_{CLS}-\Theta||^2]\\
&=E[||Z X^Te-Z H^T(HZ H^T)^{-1}(H\Theta+HZ X^Te-D)||^2]\\
&=E[||Z X^Te-Z H^T(HZ H^T)^{-1}HZ X^Te||^2]\\
&=\text{Tr}\{E[(Z X^T-Z H^T(HZ H^T)^{-1}HZ X^T)^T\\
&\ \ \ \ \cdot(Z X^T-Z H^T(HZ H^T)^{-1}HZ X^T)ee^T]\}.\\
\end{array}\nonumber
\end{equation}
From \cite{qi 2013}, we know $E(ee^T)\leq\frac{M}{4N}I$. Therefore,
\begin{equation}\label{eq30}
\begin{array}{rl}
&\ \ \ \ \text{E}[||\hat \Theta_{CLS}-\Theta||^2]\\
&\leq \frac{M}{4N}\text{Tr}[(Z X^T-Z H^T(HZ H^T)^{-1}HZ X^T)^T\\
&\ \ \ \ \cdot(Z X^T-Z H^T(HZ H^T)^{-1}HZ X^T)]\\
&=\frac{M}{4N}\text{Tr}[XZ^2X^T-XZH^T(HZ H^T)^{-1}HZ^2X^T\\
&\ \ \ \ -XZ^2H^T(HZ H^T)^{-1}HZX^T\\
&\ \ \ \ +XZH^T(HZ H^T)^{-1}HZ^2H^T(HZ H^T)^{-1}HZX^T]\\
&=\frac{M}{4N}\text{Tr}(Z)-\frac{M}{4N}\text{Tr}[(HZ H^T)^{-1}HZ^2H^T].\\
\end{array}
\end{equation}
We have
\begin{equation}
\begin{array}{rl}
HZH^T&=(I,...,I)\text{diag}[(X_0^TX_0)^{-1},...,(X_0^TX_0)^{-1}](I,...,I)^T\\
&=n(X_0^TX_0)^{-1}.\\
\end{array}\nonumber
\end{equation}
It is clear that $(HZH^T)^{-1}=X_0^TX_0/n$ and $HZ^2H^T=n(X_0^TX_0)^{-2}$. Continuing (\ref{eq30}), we have
\begin{equation}
\begin{array}{rl}
&\ \ \ \ \text{E}[||\hat \Theta_{CLS}-\Theta||^2]\\
&\leq \frac{M}{4N}\text{Tr}[I_n\otimes(X_0^TX_0)^{-1}]-\frac{M}{4N}\text{Tr}[X_0^TX_0/n\cdot n(X_0^TX_0)^{-2}]\\
&=\frac{M}{4N}n\text{Tr}[(X_0^TX_0)^{-1}]-\frac{M}{4N}\text{Tr}[(X_0^TX_0)^{-1}]\\
&=\frac{(n-1)M}{4N}\text{Tr}[(X_0^TX_0)^{-1}].\\
\end{array}\nonumber
\end{equation}
Hence, we have
\begin{equation}\label{eq28}
\begin{array}{rl}
\text{E}(\sum_i||\hat E_i-E_i||^2)&=\text{E}\{\sum_i\text{Tr}\{[\sum_{a=1}^{d^2}(\hat \theta_a^{(i)}-\theta_a^{(i)})\Omega_a]^2\}\}\\
&=\text{E}(||\hat \Theta_{CLS}-\Theta||^2)\\
&\leq\frac{(n-1)M}{4N}\text{Tr}[(X_0^TX_0)^{-1}],\\
\end{array}
\end{equation}
which we refer to as \textit{the CLS bound}.

\begin{remark}
In cases when the last POVM element $P_n$ is omitted for simplicity, unconstrained LRE can be used for stage-1 approximation, and a corresponding error upper bound can be obtained as in \cite{qi 2013}:
\begin{equation}\label{eq22}
\begin{array}{rl}
\frac{M}{4N}\text{Tr}[(X^TX)^{-1}]&=\frac{M}{4N}\text{Tr}\{[I_n\otimes(X_0^TX_0)]^{-1}\}\\
&=\frac{nM}{4N}\text{Tr}[(X_0^TX_0)]^{-1}.\\
\end{array}
\end{equation}
Comparing (\ref{eq28}) and (\ref{eq22}), we find they are only different by a factor of $\frac{n-1}{n}$. For any given detector, $n$ is fixed and these two bounds behave the same, apart from a constant. We thus omit analysis for unconstrained LRE method.
\end{remark}

\subsection{Error in $||\hat F_i-F_i||$}

We start from the spectral decomposition (\ref{eq2}). Since $\hat W_i^\dagger E_i\hat W_i$ is positive semidefinite, its diagonal elements are all nonnegative. Therefore, we have
\begin{equation}\label{eq3}
\begin{array}{rl}
&\ \ \ \ ||\hat F_i-F_i||^2=||\hat F_i-E_i||^2\\
&=\sum_{j=1}^{d-\hat n_i}[(\hat K_i)_{jj}-(\hat W_i^\dagger E_i\hat W_i)_{jj}]^2+\sum_{j= d-\hat n_i+1}^{d}(\hat W_i^\dagger E_i\hat W_i)_{jj}^2\\
&\ \ \ \ +\sum_{j=1}^{d}\sum_{k=1,k\neq j}^d|(\hat W_i^\dagger E_i\hat W_i)_{jk}|^2\\
&\leq \sum_{j=1}^{d-\hat n_i}[(\hat K_i)_{jj}-(\hat W_i^\dagger E_i\hat W_i)_{jj}]^2\\
&\ \ \ \ +\sum_{j= d-\hat n_i+1}^{d}[(\hat K_i)_{jj}-(\hat W_i^\dagger E_i\hat W_i)_{jj}]^2\\
&\ \ \ \ +\sum_{j=1}^{d}\sum_{k=1,k\neq j}^d|(\hat W_i^\dagger E_i\hat W_i)_{jk}|^2\\
&=||\hat E_i-E_i||^2.\\
\end{array}
\end{equation}

\subsection{Error in $||\hat C\hat C^\dagger-I||$}
We have the following relationship:
\begin{equation}\label{eq4}
\begin{array}{rl}
||\hat C\hat C^\dagger-I||&=||\sum_i\hat F_i-I||=||\sum_i(\hat F_i-F_i)||\\
&\leq\sum_i||\hat F_i-F_i||\leq\sum_i||\hat E_i-E_i||.\\
\end{array}
\end{equation}

\subsection{Error in $||\hat{C}\hat U-I||$}
Let $\hat S^2=\text{diag}(1+s_1,...,1+s_d)$. We have
\begin{equation}
\begin{array}{rl}
||\hat C\hat U-I||^2&=d+\text{Tr}(\hat C\hat C^\dagger)-\text{Tr}(\hat C\hat U+\hat U^\dagger\hat C^\dagger)\\
&=d+\text{Tr}(\hat C\hat C^\dagger)-2\text{Tr}(\sqrt{\hat C^\dagger\hat C})\\
&=d+\text{Tr}(\hat S^2)-2\text{Tr}(\hat S)\\
&=d+\sum_i(1+s_i)-2\sum_i\sqrt{1+s_i}\\
&=\sum_i(\sqrt{1+s_i}-1)^2\\
&=\sum_i\frac{s_i^2}{2+s_i+2\sqrt{1+s_i}}\\
&=\sum_is_i^2[\frac{1}{4}-\frac{1}{8}s_i+o(s_i)]\\
&\sim \frac{1}{4}||\hat C\hat C^\dagger-I||^2,\\
\end{array}\nonumber
\end{equation}
where the last line comes from the fact $||\hat C\hat C^\dagger-I||=||\hat U_\alpha\hat S^2\hat U_\alpha^\dagger-I||=||\hat S^2-I||=\sqrt{\sum_is_i^2}$.

Using (\ref{eq4}), we know
\begin{equation}\label{eq5}
||\hat C\hat U-I||=O(\frac{1}{2}\sum_i||\hat E_i-E_i||),
\end{equation}
where we do not incorporate the constant into the $O$ notation before the end of this proof.

\subsection{Error in $||(\hat U^\dagger\hat C^\dagger)^{-1}-I||$}

Denote the singular values of $\hat C\hat U$ as $\hat\mu_i$ for $1\leq i\leq d$. From (\ref{eq03}) we know each $\hat\mu_i^2$ is an eigenvalue of $I+\sum_i\hat G_i$. Hence, we have $\hat\mu_i\geq1$ for every $1\leq i\leq d$. Therefore,
\begin{equation}\label{eq31}
||(\hat U^\dagger\hat C^\dagger)^{-1}||=\sqrt{\sum_i\frac{1}{\hat \mu_i^2}}\leq {\sqrt{d}}.
\end{equation}

Using (\ref{eq5}), we have
\begin{equation}\label{eq32}
\begin{array}{rl}
&\ \ \ \ ||(\hat U^\dagger\hat C^\dagger)^{-1}-I||\\
&=||(\hat U^\dagger\hat C^\dagger)^{-1}(\hat U^\dagger\hat C^\dagger-I)^2-(\hat U^\dagger\hat C^\dagger-I)||\\
&\leq||(\hat U^\dagger\hat C^\dagger)^{-1}||\cdot ||\hat U^\dagger\hat C^\dagger-I||^2+||\hat U^\dagger\hat C^\dagger-I||\\
&\leq {\sqrt{d}}||\hat U^\dagger\hat C^\dagger-I||^2+||\hat U^\dagger\hat C^\dagger-I||\\
&\sim ||\hat U^\dagger\hat C^\dagger-I||=||\hat{C}\hat U-I||\\
&=O(\frac{1}{2}\sum_i||\hat E_i-E_i||),\\
\end{array}
\end{equation}
where the last $\sim$ comes from omitting the higher order term.

\subsection{Error in $\sum_i||\hat P_i-P_i||^2$}
Since each $F_i=E_i$ is positive semidefinite, we have
\begin{equation}
\begin{array}{rl}
 \sum_i||F_i||^2&=\sum_i\text{Tr}(E_i^2)=\text{Tr}(\sum_iE_i^2)\leq\text{Tr}(\sum_iE_i^2+\sum_{i,j}E_iE_j)\\
 &=\text{Tr}[(\sum_iE_i)^2]=\text{Tr}(I)=d.\\
\end{array}\nonumber
\end{equation}

For each $i$, we have
$$||F_i||=||E_i||\leq ||I||=\sqrt{d}.$$

Using (\ref{eq3}), (\ref{eq5}), (\ref{eq31}) and (\ref{eq32}), we have
\begin{equation}\label{eqa30}
\begin{array}{rl}
&\ \ \ \ \sum_i||\hat P_i-P_i||^2\\
&=\sum_i||(\hat C\hat U)^{-1}\hat F_i(\hat U^\dagger\hat C^\dagger)^{-1}-F_i||^2\\
&=\sum_i||(\hat C\hat U)^{-1}\hat F_i(\hat U^\dagger\hat C^\dagger)^{-1}-\hat F_i(\hat U^\dagger\hat C^\dagger)^{-1}\\
&\ \ \ \ +\hat F_i(\hat U^\dagger\hat C^\dagger)^{-1}-\hat F_i+\hat F_i -F_i||^2\\
&\leq\sum_i[||(\hat C\hat U)^{-1}\hat F_i(\hat U^\dagger\hat C^\dagger)^{-1}-\hat F_i(\hat U^\dagger\hat C^\dagger)^{-1}||\\
&\ \ \ \ +||\hat F_i(\hat U^\dagger\hat C^\dagger)^{-1}-\hat F_i||+||\hat F_i -F_i||]^2\\
&\leq\sum_i[||(\hat C\hat U)^{-1}-I||\cdot||\hat F_i(\hat U^\dagger\hat C^\dagger)^{-1}||\\
&\ \ \ \ +||\hat F_i||\cdot||(\hat U^\dagger\hat C^\dagger)^{-1}-I||+||\hat F_i -F_i||]^2\\
&\leq\sum_i[||(\hat U^\dagger\hat C^\dagger)^{-1}-I||\cdot||\hat F_i[(\hat U^\dagger\hat C^\dagger)^{-1}-I]||\\
&\ \ \ \ +||(\hat U^\dagger\hat C^\dagger)^{-1}-I||\cdot||\hat F_i\cdot I||\\
&\ \ \ \ +||\hat F_i||\cdot||(\hat U^\dagger\hat C^\dagger)^{-1}-I||+||\hat F_i -F_i||]^2\\
&\sim \sum_i[2||\hat F_i||\cdot||(\hat U^\dagger\hat C^\dagger)^{-1}-I||+||\hat F_i -F_i||]^2\\
&\leq \sum_i[||\hat F_i||O(\sum_j||\hat E_j-E_j||)+||\hat E_i -E_i||]^2\\
&=\sum_i[||\hat F_i||^2O(\sum_j||\hat E_j-E_j||)^2+||\hat E_i -E_i||^2\\
&\ \ \ \ +2||\hat F_i||O(\sum_j||\hat E_j-E_j||)||\hat E_i -E_i||]\\
&\sim O(\sum_j||\hat E_j-E_j||)^2\sum_i||F_i||^2+\sum_i||\hat E_i -E_i||^2\\
&\ \ \ \ +2O(\sum_j||\hat E_j-E_j||)\sum_i||F_i||\cdot||\hat E_i -E_i||\\
&\leq d\cdot O(\sum_j||\hat E_j-E_j||)^2+\sum_i||\hat E_i -E_i||^2\\
&\ \ \ \ +2\sqrt{d}O(\sum_j||\hat E_j-E_j||)\sum_i||\hat E_i -E_i||\\
&\leq dn\cdot O(\sum_j||\hat E_j-E_j||^2)+\sum_i||\hat E_i -E_i||^2\\
&\ \ \ \ +2\sqrt{d}nO(\sum_j||\hat E_j-E_j||^2)\\
&=(dn+2\sqrt{d}n+1)O(\sum_j||\hat E_i-E_i||^2),\\
\end{array}
\end{equation}
where the first $\sim$ results from omitting the higher order terms, and the second $\sim$ comes from the relationship $||\hat{F}_i||\leq||\hat{F}_i-F_i||+||F_i||\sim||F_i||$, and the last inequality comes from the Cauchy-Schwarz inequality
$$(\sum_i||\hat E_i-E_i||)^2\leq n(\sum_i||\hat E_i-E_i||^2).$$

Taking the expectation of (\ref{eqa30}) and using (\ref{eq28}), we have
\begin{equation}\label{bound1}
\begin{array}{rl}
&\ \ \ \ \text{E}(\sum_i||\hat P_i-P_i||^2)\\
&= (dn+2\sqrt{d}n+1)O[\text{E}({\sum_i||\hat E_i-E_i||^2})]\\
&= O\{\frac{(dn+2\sqrt{d}n+1)(n-1)M}{4N}\text{Tr}[(X_0^TX_0)^{-1}]\}.\\
\end{array}
\end{equation}
Since we have explicitly shown the constants in the $O$ notation, (\ref{bound1}) should be interpreted as that the following equation holds asymptotically:
\begin{equation}\label{eqb5}
\begin{array}{rl}
&\ \ \ \ \text{E}(\sum_i||\hat P_i-P_i||^2)\\
&\leq \frac{(dn+2\sqrt{d}n+1)(n-1)M}{4N}\text{Tr}[(X_0^TX_0)^{-1}]+o(\frac{1}{N}).\\
\end{array}
\end{equation}
Using (\ref{assum2}), we can further simplify (\ref{bound1}) as
\begin{equation}\label{eqb6}
\text{E}(\sum_i||\hat P_i-P_i||^2)= O(\frac{d^5n^2}{N}).
\end{equation}
\end{IEEEproof}

\begin{remark}
If the probe states are not optimal, (\ref{eqb6}) might fail and only (\ref{eqb5}) holds. This proof also indicates that $\text{Tr}[(X_0^TX_0)^{-1}]$ is a helpful index to guide the choice of the probe states. If different probe states are highly similar to each other, then they result in a large $\text{Tr}[(X_0^TX_0)^{-1}]$ and thus a large estimation error.
\end{remark}

\section{Optimization of the Coherent Probe States}\label{sec5}
Since the detector to be estimated is usually unknown in practice, the optimization among all the possible probe states should be independent of the specific detector. An advantage of our TSE method is that an explicit error upper bound is presented, which does not involve the specific form of the detector. This can be critical in the optimization of the probe states. Moreover, to adapt to practical applications, we assume the probe states are all coherent states in this section.

\subsection{On the types of probe states}\label{subsec5}

In quantum optics experiments, the preparation of number states $|k\rangle$ ($k\in\mathbb{N}$) is a difficult task, especially when $k$ is large. Therefore, in practice the input probe states are usually coherent states instead. A coherent state is denoted as $|\alpha\rangle$ where $\alpha\in\mathbb{C}$ and it can be expanded using number states as
$$|\alpha\rangle=\text{e}^{-\frac{|\alpha|^2}{2}}\sum_{i=0}^{\infty}\frac{\alpha^i}{\sqrt{i!}}|i\rangle.$$
Their inner product relationship is
\begin{equation}\label{eq39}
\langle\beta|\alpha\rangle=\text{e}^{-\frac{1}{2}(|\beta|^2+|\alpha|^2-2\beta^*\alpha)}.
\end{equation}
We usually identify $\alpha$ with $|\alpha\rangle$ when there is no ambiguity.

Let $|\alpha\rangle_d=\text{e}^{-\frac{|\alpha|^2}{2}}\sum_{i=0}^{d-1}\frac{\alpha^i}{\sqrt{i!}}|i\rangle$. Coherent states are in essence infinite dimensional. To estimate a $d$-dimensional detector, in practice we employ $|\alpha\rangle_d$ as the (approximate) mathematical description of $|\alpha\rangle$ in this paper. Throughout this paper we assume the detector gives no signal when saturated, which means the part $|\alpha\rangle-|\alpha\rangle_d$ can be distinguished from $|\alpha\rangle_d$.

The tail part $|\alpha\rangle-|\alpha\rangle_d$ can be viewed as noise, which should be suppressed. This requires the amplitude $|\alpha|$ to be not large. Furthermore, (\ref{eq39}) indicates that if $|\alpha|$ and $|\beta|$ are both close to zero, their inner product will also be close to one, which means that coherent states with small amplitudes are very much ``alike". This indicates that we cannot employ probe state sets where all the amplitudes are small. Considering the above two requirements, we design the preparation procedure of the probe states as follows.

\textit{Probe States Preparation:} Given appropriate $q>0$, generate two random numbers $x$ and $y$ independently with their probability density function uniformly distributed on $[-q,q]$. Then $|x+\text{i}y\rangle$ will be employed as a probe state, with $N/M$ copies. Repeat this process to generate $M$ probe states and employ them to perform detector tomography.

\begin{remark}
Our sampling procedure is in essence sampling randomly within a given square in the complex plane. Another candidate method is to sample following a certain symmetric fixed pattern within this given square. Since simulation shows little difference in the final estimation error, we stick to our random-sample procedure.
\end{remark}

With our probe state preparation procedure, we wonder what is the relationship between $M$ and the final estimation error, when other factors, such as the detector, the total number of copies $N$ for the probe states and the parameter $q$, remain unchanged.

To ensure that the inversion of $X^TX$ in (\ref{eqa09}) exists, it is required that at least $M\geq d^2$. We further find that when $M$ is large enough, the final estimation error tends to a constant independent of $M$. We give an explanation as follows.

First, the $j$th probe state $|\alpha\rangle$ is approximately viewed as $|\alpha\rangle_d$, which has a corresponding parametrization $\Phi_j$. Let $\mathbb{E}(\cdot)$ denote the expectation of functions of $x$ and $y$, in contrast to the expectation $\text{E}(\cdot)$ in Theorem \ref{the1}. Let $f_j=\Phi_j-\mathbb{E}(\Phi_j)$. Then the $f_j$s are i.i.d. with respect to the subscript $j$. According to (\ref{eq28}), the estimation error upper bound is $\frac{(n-1)M}{4N}\text{Tr}[(X_0^TX_0)^{-1}]$. We thus have
\begin{equation}\label{eq40}
\begin{array}{rl}
&\ \ \ \ \mathbb{E}\{\frac{(n-1)M}{4N}\text{Tr}[(X_0^TX_0)^{-1}]\}\\
&=\frac{n-1}{4N}\text{Tr}\{M\mathbb{E}[(\sum_{j=1}^M (\mathbb{E}(\Phi_j)+f_j)(\mathbb{E}(\Phi_j)^T+f_j^T))^{-1}]\}\\
&=\frac{n-1}{4N}\text{Tr}\{M\mathbb{E}[(M\mathbb{E}(\Phi_j)\mathbb{E}(\Phi_j)^T+\sum_{j=1}^M f_jf_j^T)^{-1}]\}\\
&=\frac{n-1}{4N}\text{Tr}\{[\mathbb{E}(\Phi_j)\mathbb{E}(\Phi_j)^T+\mathbb{E}(\frac{1}{M}\sum_{j=1}^M f_jf_j^T)]^{-1}\}.\\
\end{array}
\end{equation}
According to the central limit theorem \cite{chow1997}, as $M$ tends to infinity, $\mathbb{E}(\frac{1}{M}\sum_{j=1}^M f_jf_j^T)$ converges to a fixed matrix, and hence the expectation of estimation error tends to a constant.

Two points should be noted: (i) In practice $M$ cannot be arbitrarily large when $N$ is given. (ii) There is usually a gap between this bound (\ref{eq40}) and the practical error. However, simulation results imply the effectiveness of the above analysis, which suggests that a modest number of different types of probe states should be enough for practical applications. To investigate the least $M$ that suffices for an estimation task with a given dimension, it only requires us to calculate $\mathbb{E}\{\frac{(n-1)M}{4N}\text{Tr}[(X_0^TX_0)^{-1}]\}$ for several candidates of $M$, which is a quantity independent of the specific detector.

\subsection{Optimization of the size of sampling square for probe states}\label{subsec6}
As analyzed in Sec. \ref{subsec5}, the estimation error would be large if $q$ is too small or too large. Hence, there should be an optimal value for the choice of $q$. This is further verified by the simulation results in Fig. \ref{test3}.

To locate the optimal value of $q$, we consider the projection of a probe state onto the $d$-dimensional subspace where the detector resides. Theoretically, the optimal value of $q$ should be different for different detectors, even though the dimension is fixed. However, in simulations (for example, Fig. \ref{test3}), we find that the optimal values for a practical detector and the bound $\mathbb{E}\{\frac{(n-1)M}{4N}\text{Tr}[(X_0^TX_0)^{-1}]\}$ usually coincide. Therefore, as an approximation, we can investigate the optimization of this bound w.r.t. $q$. Furthermore, the value of $N$ does not affect this optimization, and from Sec. \ref{subsec5} we know an $M$ not too small will also be irrelevant to the optimization. Hence, we only need to optimize
\begin{equation}\label{eq41}
\mathbb{E}\{\text{Tr}[(X_0^TX_0)^{-1}]\},
\end{equation}
which is a quantity uniquely determined by the probe states.

We start from the real function defined on all nonnegative integers $g(k)=\text{e}^{-|\alpha|^2}\frac{|\alpha|^{2k}}{k!}$, where $\alpha$ is the corresponding complex number of a probe state $|\alpha\rangle$. From $g(k)-g(k+1)=\text{e}^{-|\alpha|^2}|\alpha|^{2k}\frac{k+1-|\alpha|^2}{(k+1)!}$, we know $g(k)$ first increases and then decreases after $k\geq|\alpha|^2-1$. Hence, $g(k)$ reaches the maximum value around $|\alpha|^2-1$.

\begin{figure}
\centering
\includegraphics[width=3.6in]{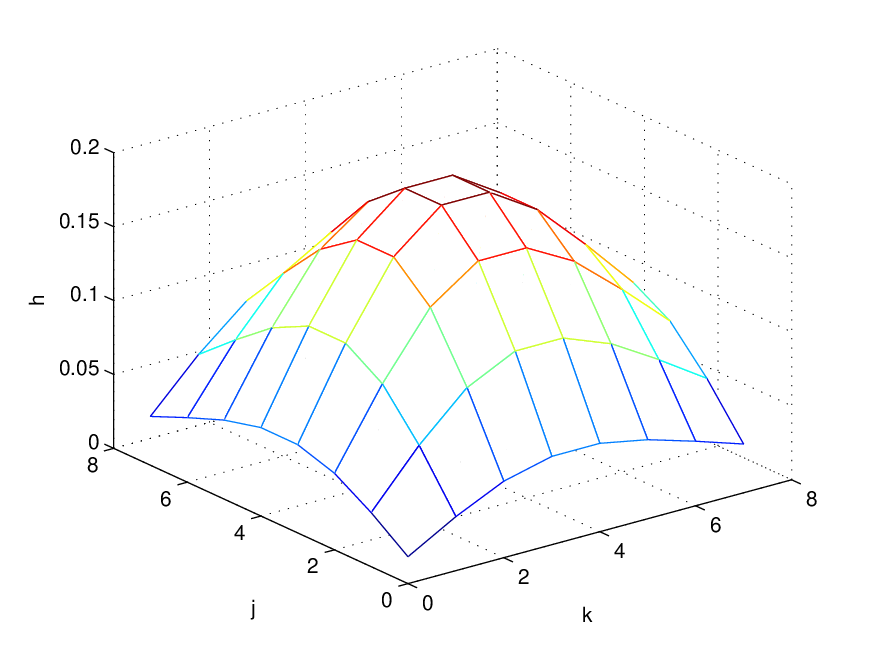}
\centering{\caption{The projection amplitude function $h(k,j)$.}\label{test00}}
\end{figure}

For any given probe state $|\alpha\rangle$, we consider the amplitude of its projection on each position $|k\rangle\langle j|$, which is
$$h(k,j)\triangleq|\langle j|\alpha\rangle\langle\alpha|k\rangle|=\text{e}^{-|\alpha|^2}\frac{|\alpha|^{j+k}}{\sqrt{j!k!}}.$$
Fig. \ref{test00} shows the grided $h(k,j)$ with $d=8$ and $|\alpha|=2$. Note that $g(k)$ is the restriction of $h(k,j)$ on $j=k$. Using the same technique for analyzing $g(k)$, it is straightforward to prove that grided $h(k,j)$ always has a single peak, with the position of the maximum around $(|\alpha|^2-1,|\alpha|^2-1)$. Generally, the larger $h(k,j)$ is, the better accuracy one can expect to obtain for estimating the element of a detector at position $|k\rangle\langle j|$. To obtain the least estimation error, a natural idea is to maximize $h(k,j)$ for each position $(k,j)$. However, this is not practical, because from $\sum_{k=0}^\infty g(k)=1$ one can see that $\sum_{k,j}h(k,j)$ is bounded. Therefore, to locate the optimal $q$ means to optimally allocate $h(k,j)$ on the $d\times d$ positions.

Generally speaking, when estimating a multivariate target $\{\theta_i\}$, the MSE $E(\sum_i|\theta_i-\hat\theta_i|^2)$ is usually dominated by the worst estimated parameter $\max_i|\theta_i-\hat\theta_i|$. Hence, the optimal $q$ (denoted as $q_o$) should have a good performance for the worst estimation. When $q$ is too small, $\mathbb{E}(|\alpha\rangle\langle\alpha|)$ is overly concentrated near the original point, and the projections on $(k,j)$s far from the original point will be too small, resulting in a large bound in (\ref{eq41}); i.e., a bad estimate. Conversely, $q$ should not be too large. If we approximately view $h(k,j)$ as symmetric, it is natural to conclude that the projection of the maximum (or the middle point of the two maxima) of $h(k,j)$ should be at $(\frac{d-1}{2},\frac{d-1}{2})$ for $q_o$. More specifically, if $|\alpha|^2-1$ is an integer, then $g(|\alpha|^2-1)=g(|\alpha|^2)$ are the two maxima. When $d$ is even, the maximum of $h(k,j)$ should be two contour points $(\frac{d-2}{2},\frac{d-2}{2})$ and $(\frac{d}{2},\frac{d}{2})$, and we should have $\frac{d-2}{2}=|\alpha|^2-1$. When $d$ is odd, $h(k,j)$ has one maximum and its projection should be $(\frac{d-1}{2},\frac{d-1}{2})$, which further indicates $\frac{d-1}{2}=\frac{|\alpha|^2+|\alpha|^2-1}{2}$. Therefore for $q_o$, we should always have
\begin{equation}\label{eq42}
(\mathbb{E}|\alpha|)^2=\frac{d}{2}.
\end{equation}
From our sampling scheme for probe states in Sec. \ref{subsec5}, we have
\begin{equation}\label{eq43}
\mathbb{E}|\alpha|=\int_{-q}^{q}\int_{-q}^{q}\frac{\sqrt{x^2+y^2}}{q^2}\,dxdy=\frac{\sqrt{2}+\ln(1+\sqrt{2})}{3}q.
\end{equation}
Combining (\ref{eq42}) and (\ref{eq43}), we have the following heuristic formula
\begin{equation}\label{eq44}
q_o=\frac{3\sqrt{d}}{2+\sqrt{2}\ln(1+\sqrt{2})}.
\end{equation}

\begin{remark}
If the probe state is the tensor product of single-qubit probe states, then one only needs to optimize each single-qubit probe state, which corresponds to the 2-dimensional edition of (\ref{eq41}). This can be straightforwardly achieved by running a numerical simulation, and the result is also covered in Fig. \ref{test4}.
\end{remark}

\section{Numerical Results}\label{secsimu}

\subsection{Basic performance}\label{simu1}
\begin{figure}
\centering
\includegraphics[width=3.6in]{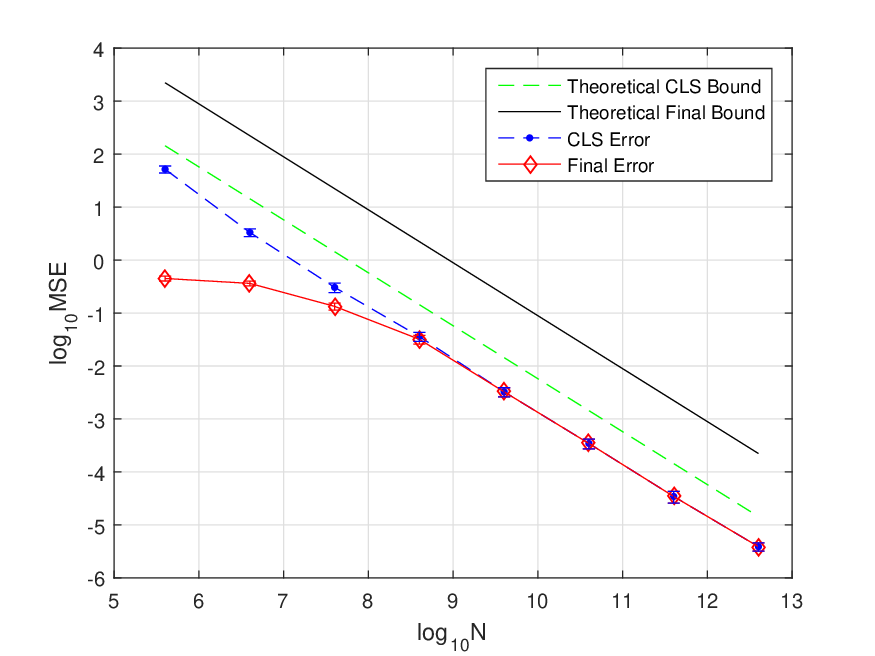}
\centering{\caption{MSE versus the logarithm of the total resource number $N$.}\label{test0}}
\end{figure}

We simulate the estimation error under different total resource numbers. We consider a $2$-dimensional system with a detector $P_1=\left(\begin{matrix}
0&0\\
0&0.3\\
\end{matrix}\right)$, $P_2=\left(\begin{matrix}
0.1&-0.02i\\
0.02i&0.2\\
\end{matrix}\right)$ and $P_3=\left(\begin{matrix}
0.9&0.02i\\
-0.02i&0.5\\
\end{matrix}\right)$. The sampling parameter for coherent states is $q=0.015$. The number of different types of probe states is $M=40$. We employ our method to estimate the detector using different resource numbers and present the results in Fig. \ref{test0}. In Fig. \ref{test0}, the green dashed line is the theoretical CLS error upper bound (\ref{eq28}), the black line is the theoretical final error upper bound (\ref{bound1}) (or equivalently, (\ref{eqb5}) without the higher order term), and the blue dots and red diamonds are the CLS error and final error, respectively. The horizontal axis is the logarithm of the total number of copies of probe states $N$ and the vertical axis is the logarithm of the Mean Square Error (MSE) $\text{E}(\sum_i||\hat P_i-P_i||^2)$. Each point in Fig. \ref{test0} is the average of 50 simulations.

In Fig. \ref{test0}, the CLS bound is better than the final bound, which is because more relaxation procedures are used to deduce the final bound and make it looser. When the resource number is large ($N>10^7$), the decreasing slope is close to $-1$, which verifies Theorem \ref{the1}. We also notice that when the resource number is small ($N<10^6$), the final estimation error is notably better than the CLS error. This is because the estimation error of arbitrary physical estimation is in essence bounded by a constant, while the CLS estimation can be nonphysical and thus leads to an unbounded error. As a result, when the resource number is not large enough, the CLS estimation is rough and the error exceeds this constant, while the final error is still bounded by this constant. This phenomenon disappears if $q$ is instead set close to the optimal value, because the final error will be too small to be influenced by the constant bound. For example, if $q=1.307$ as predicted by (\ref{eq44}), the MSEs in Fig. \ref{test0} will decrease by $6$ orders of magnitude.

\subsection{On the types of probe states}\label{simu2}

\begin{figure}
\centering
\includegraphics[width=3.6in]{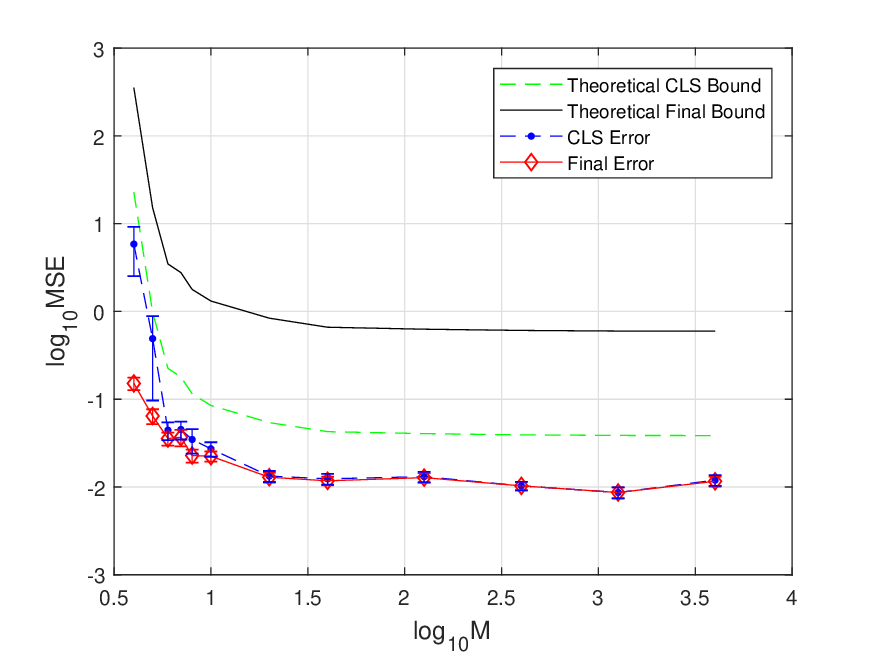}
\centering{\caption{MSE versus probe state types $M$.}\label{test1}}
\end{figure}

We simulate the performance of our algorithm with different number of types of probe states. The detector and $q$ are the same as in Sec. \ref{simu1}. The total resource number is $1.44\times 10^{9}$.  We perform our estimation method with $M$ varying from $4$ to $4000$, and present the results in Fig. \ref{test1}, where each point is the average of 100 simulations. The legend is the same as Fig. \ref{test0}, except that the horizontal axis is the number of types of probe states $M$ in logarithm. We can see when $M$ is very small, both the theoretical bound and the practical errors are large, due to the fact that the probe states lack diversity and their linear dependence is high. When $M$ is over $10$, both the bound and the practical errors quickly tend to constants, which validates our analysis in Sec. \ref{subsec5}. Therefore, in practice a moderate number of different probe states should suffice.

\subsection{Optimization of the size of sampling square for probe states}\label{simu3}

\begin{figure}
\centering
\includegraphics[width=3.6in]{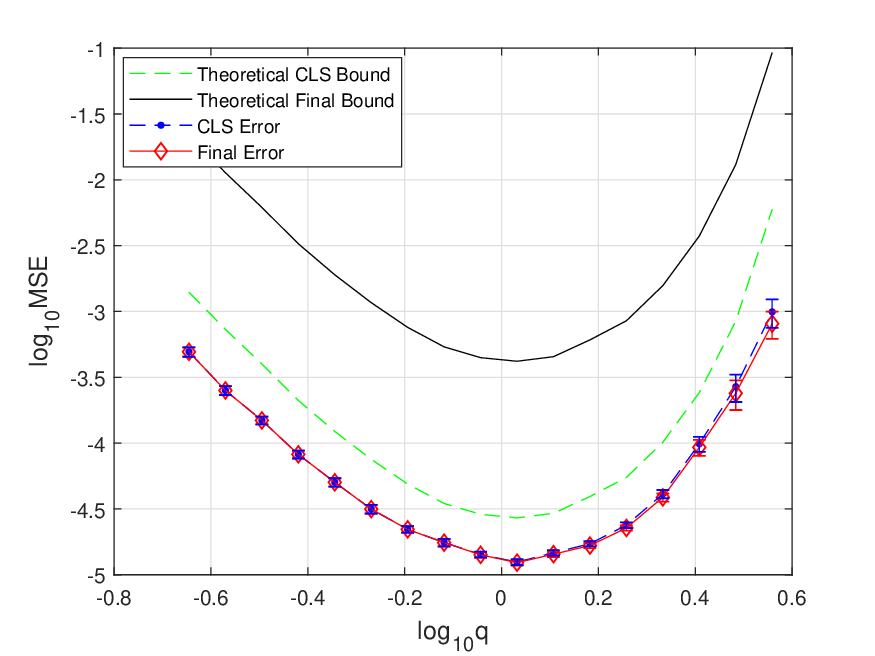}
\centering{\caption{MSE versus the size $q$ of sampling square for probe states.}\label{test3}}
\end{figure}

We perform simulations to illustrate that the optimal size of the sampling square for coherent probe states coincides with the optimal point of the bound (\ref{eq41}). We consider a system with the same detector as that in Sec. \ref{simu1}. The total resource number is $N=10^{6}$, and the number of different types of probe states is $M=32$. We perform our estimation method under different $p$, and present the results in Fig. \ref{test3}. Each point is the average of 200 simulations. We can see that there is indeed an optimal point for the practical estimation error with respect to different sizes $q$, which validates the analysis in Sec. \ref{subsec6}. Also this practical optimal position of $q$ basically coincides with the optimal position of the error bound.

\begin{figure}
\centering
\includegraphics[width=3.6in]{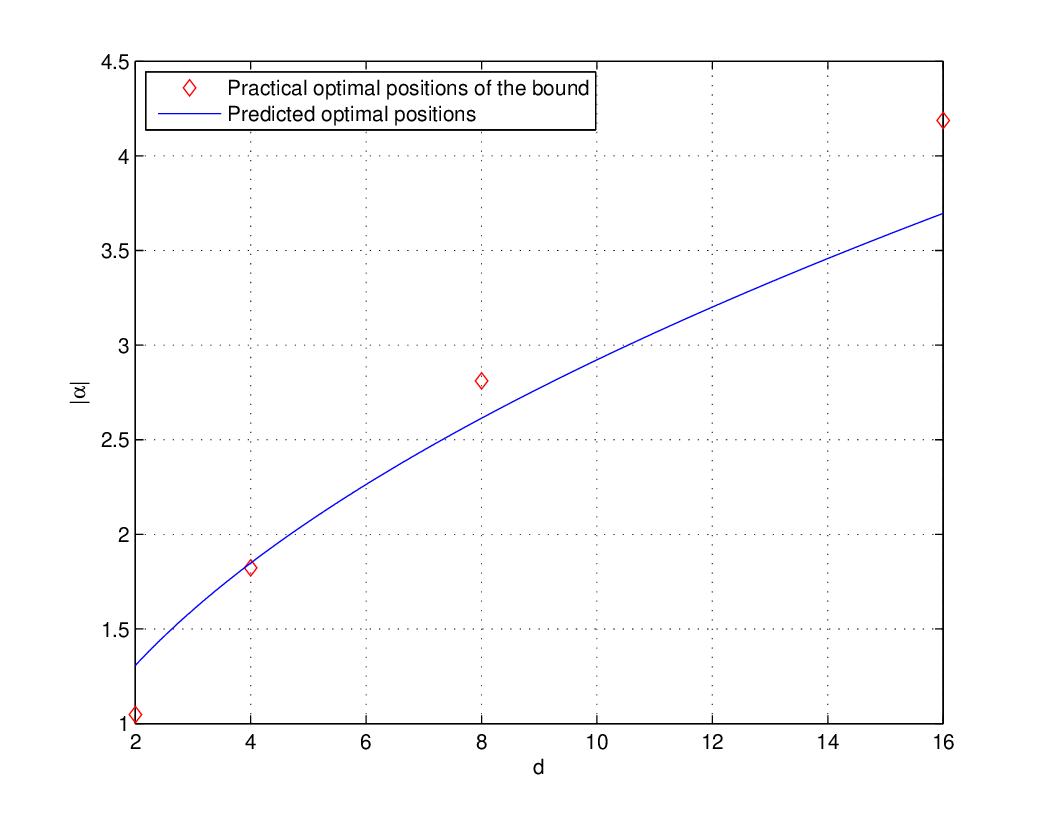}
\centering{\caption{The optimal sampling square size $q_o$ versus dimension $d$.}\label{test4}}
\end{figure}

Using the same system we simulate to search for the optimal size $q$ of the sampling square for probe states in different dimensions. The practical optimal positions we search for are the minimum points of the bound (\ref{eq41}) under dimensions $d=2,4,8,16$, which are presented as red diamonds in Fig. \ref{test4}. The blue line is the optimal $q_o$ predicted by our formula (\ref{eq44}), which are close to the practical optimal values still with improvement space.

\subsection{Comparison with MLE using qubit probes}

\begin{figure}
\centering
\includegraphics[width=3.0in]{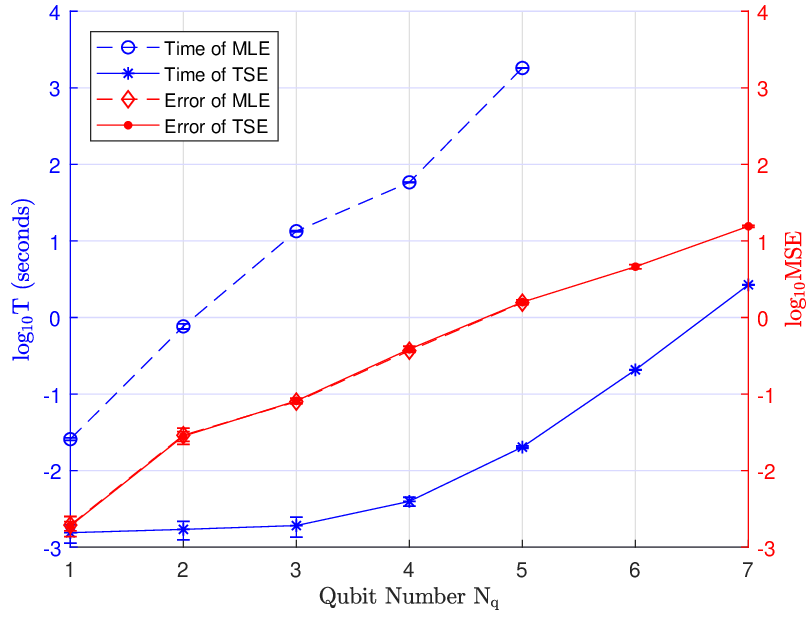}
\centering{\caption{Comparison between our TSE algorithm with MLE for different qubit number $N_q$.}\label{testmle}}
\end{figure}

\begin{figure}
\centering
\includegraphics[width=3.3in]{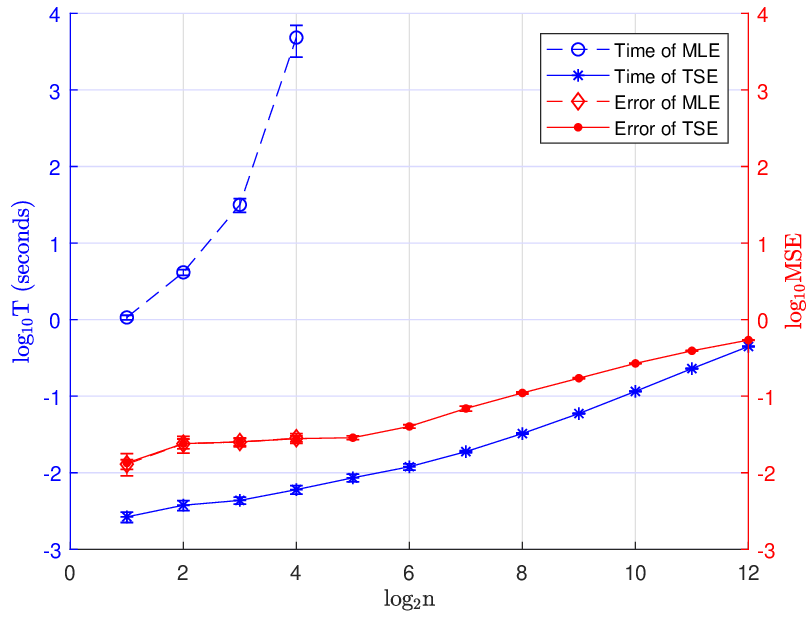}
\centering{\caption{Comparison between our algorithm with MLE for different $n$.}\label{testn}}
\end{figure}

We compare our TSE method with the Maximum Likelihood Estimation (MLE) method, which is one of the most widely used methods. We simulate an $N_q$-qubit detector with $P_1+P_2=I$ where $$P_1=U_1\text{diag}(1,\frac{1}{2},\frac{1}{3},...,\frac{1}{2^{N_q}})U_1^\dagger$$ and $$U_1=\left[\frac{1}{{2}}\left(\begin{matrix}
1 & \sqrt{3}\\
-\sqrt{3} & 1\\
\end{matrix}\right)\right]^{\otimes N_q}.$$

The probe states are the tensor product of single-qubit states $\{\frac{I}{2},\frac{I+\sigma_x}{2},\frac{I+\sigma_y}{2},\frac{I+\sigma_z}{2}\}$ where $$\sigma_x=\left(\begin{matrix}
0&1\\
1&0\\
\end{matrix}\right),\sigma_y=\left(\begin{matrix}
0&-i\\
i&0\\
\end{matrix}\right),\sigma_z=\left(\begin{matrix}
1&0\\
0&-1\\
\end{matrix}\right).$$ For each $N_q$, the total resource number of the probe states is $N=10^3\times 2^{3N_q}$, and they are evenly distributed to each probe state. The MLE algorithm we use is the method in \cite{paris 2004}. We compare the estimation results of our TSE method and MLE in Fig. \ref{testmle}, where each point is the average of 10 simulations. The running time ($T$) is the online computational time. Note that the detector tomography method via MLE is in essence a numerical search algorithm and a theoretical characterization of the computational complexity is challenging. For each detector, we first run our algorithm, and then adjust the MLE method such that the averaged estimation error of MLE is within $[95\%,105\%]$ of the error of our algorithm. We see that for $N_q\geq 4$ qubits our algorithm can be faster than MLE by over 4 orders of magnitude. In this simulation, $M=d^2$, and we thus anticipate the computational complexity is $O(d^4)$, which indicates a theoretical slope $1.204$ for our running time in the coordinate of Fig. \ref{testmle}. For the simulated running time of our algorithm, the slope of the fitting line of the right three points is $1.060$, which is close to the theoretical value but still with some difference, possibly because the qubit number is not large enough.

It is difficult to rigorously compare the computational complexity of the two algorithms in some averaging sense of all possible detectors. To give a simple illustration, we fix $N_q=3$ and $N=10^3\times 2^9$, and change the detector as $P_1'+P_2'=I$ with $P_1'=kU_RP_1U_R^\dagger$, where $k$ is a random variable evenly distributed in $(0,1)$ and $U_R$ is a random unitary matrix generated following the algorithm in \cite{zyczkowski94}. We independently generate 10 pairs of $k$ and $U_R$ and compare the averaged performance of our algorithm and MLE for these 10 pairs. For each pair, we still first run TSE method (with 10 repetitions) and then adjust MLE so that the averaged estimation error of MLE is within $[95\%,105\%]$ of the error of TSE. The final 10-pair-averaged error of TSE is $0.0581\pm 4.58\times 10^{-3}$, and $0.0584\pm 4.64\times10^{-3}$ for MLE. The 10-pair-averaged running time of TSE is $1.50\times10^{-3}\pm8.14\times10^{-5}$, and $9.39\pm1.08$ for MLE, in seconds. This result generally matches the performance in Fig. \ref{testmle}.

We also simulate the case when $n$ increases. We fix $d=4$ and the detector is
\begin{equation}
P_j=V_j\text{diag}[\frac{1}{n},(1,\frac{1}{2},\frac{1}{3})\frac{j}{n^2}]V_j^\dagger,
\end{equation}
where for $j<n$ we have $$V_j=\left\{
\begin{array}{ll}
\frac{1}{\sqrt{2}}\left(\begin{matrix}
1&1\\
1&-1\\
\end{matrix}\right)\otimes\frac{1}{\sqrt{2}}\left(\begin{matrix}
1&1\\
1&-1\\
\end{matrix}\right)^\dagger, & \text{when j is odd.}\\
e^{-\text{i}\sigma_x\otimes \sigma_x}, &\text{when j is even.}\\
\end{array}
\right.$$
The probe states are the same as those in the above simulation. We choose $n$ to be a power of 2 and run the simulation for different values of $n$. The total resource number of the probe states is fixed as $N=10^3\times 2^{3}$, and they are evenly distributed to each probe state. We plot the running time ($T$) versus $n$ in logarithmic coordinates for our TSE method and MLE in Fig. \ref{testn}, where each point is the average of 10 simulations. We see that TSE can be significantly faster than MLE for large $n$. Theoretically, $T=O(n)$ indicates a slope $0.301$ for our method. For the simulated running time of our algorithm, the slope of the fitting line of the right three points is $0.293$, which is close to the theoretical value. Furthermore, Fig. \ref{testmle} and Fig. \ref{testn} also imply the relationship between the estimation error and $n$ and $d$, which is not close to the prediction of (\ref{eqb6}). One possible reason is that the bound (\ref{eqb6}) might not be tight. Also, note that the practical error is dependent on the specific detector and when $n$ and $d$ change the detector necessarily changes. Hence, we leave it an open problem to better characterize the increasing tendency of the error w.r.t. $n$ and $d$.

\begin{remark}
For practical detectors $n$ is usually smaller than $d$. However, this pattern means a very large $d$ in simulation, which is difficult to perform if we are to simulate the performance of MLE as comparison. Hence, we do not enforce large $d$ when performing simulation in Fig. \ref{testn}.
\end{remark}

\section{Experimental Results}\label{secexp}

\subsection{Experimental setup}\label{expset}
\begin{figure}
	\includegraphics[width=3.5in]{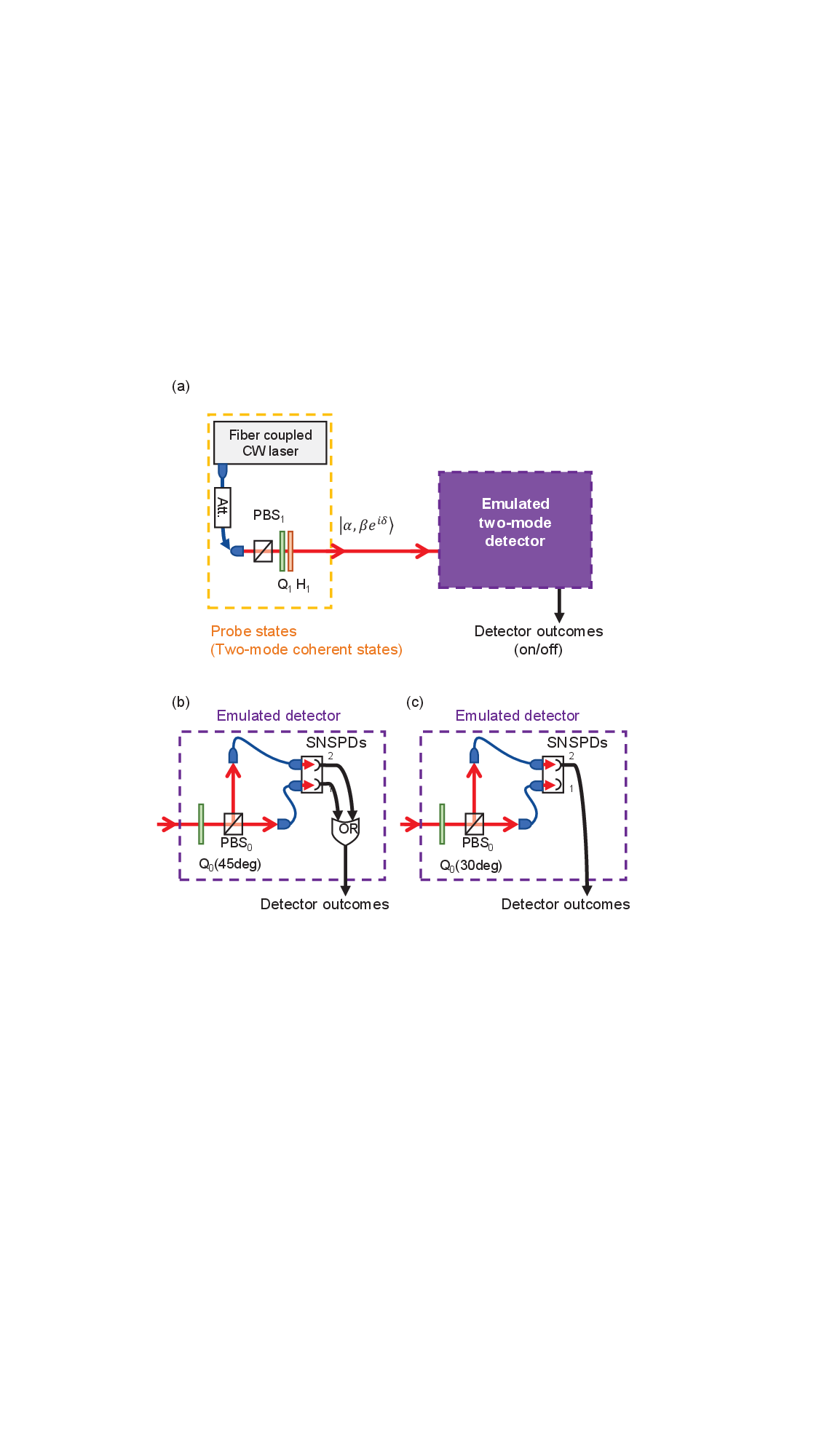}
	\caption{
		 Experimental setup \cite{shota 2017}.
			Att., Attenuator;
			PBS, Polarization beam splitter;
			H, Half wave plate;
			Q, Quarter wave plate;
			SNSPD, Superconducting nanowire single photon detector.
	}
	\label{fig}
\end{figure}

We first briefly explain the entire experimental setup (as in Fig. \ref{fig}(a)), which determines the structure of the detector to be estimated. More details about this setup can be found in \cite{shota 2017}.

In Fig. \ref{fig}, the purple dashed box corresponds to the emulated quantum detector which works as two-mode inputs - one binary output detector.
Two independent quantum modes are encoded within orthogonal polarization modes in one optical beam at the detector input.
The two-mode quantum detector consists of two superconducting nanowire detectors (SNSPDs), a polarization beam splitter (PBS), a quarter wave plate (QWP), and a logical OR gate.
The polarization of the input beam is first rotated by a QWP$_0$ with the azimuth angle of 45$^\circ$ (Fig. \ref{fig}(b)), or 30$^\circ$ (Fig. \ref{fig}(c)), respectively. Then the beam is split into two spatially separated beams via PBS$_0$, and they are injected into two SNSPDs through optical fibers.
The photon counting signals from the two SNSPDs are sent to a logical OR gate, and the final detector output is obtained as on/off signal corresponding to POVMs of $P_1$ and $P_0$ ($P_0 + P_1 = I$). Fig. \ref{fig}(b) and (c) are different specific settings to generate different emulated detectors.

This experimental setup leads to a special class of detectors. Specifically, we require them to be block diagonal (e.g., see \cite{shota 2017}):
\begin{equation}\label{eqa43}
P_i=L^{(i)}_1\oplus L^{(i)}_2\oplus ...\oplus L^{(i)}_m,
\end{equation}
where $m$ is the number of different blocks and $L^{(i)}_j\geq 0$ is $d_j\times d_j$ dimensional, with $\sum_{j=1}^m d_j=d$. Hence, we need to modify our original TSE method to reconstruct $\{P_i\}$.

\subsection{Modified TSE protocol}

First we choose $\{\Omega_i\}_{i=1}^v$ to be a complete orthogonal Hermitian basis set for the space of $\{P_i\}$ (instead of for $\mathbb{C}_{d\times d}$), where $\Omega_1=I_d/\sqrt{d}$ and $v$ equals to ${\sum_j d_j^2}$ instead of $d^2$. Then we have the parametrization under this basis set as
$$P_i=\sum_{a=1}^{v}\theta_a^{(i)}\Omega_a,$$
$$\rho_j=\sum_{b=1}^{v}\phi_b^{(j)}\Omega_b,$$
and the theoretical probability is $p_{ij}=\text{Tr}(P_i\rho_j)$, which now becomes
$$p_{ij}=\sum_{a=1}^{v}\phi_a^{(j)}\theta_a^{(i)}\triangleq \Phi_j^T\Theta_i.$$
The linear regression equation is now
$$\hat p_{ij}=\Phi_j^T\Theta_i+e_{ij},$$
and the error $e_{ij}=\hat p_{ij}- p_{ij}$ converges in distribution to a normal distribution with mean 0 and variance $(p_{ij}-p_{ij}^2)/(N/M)$. Let $\Theta=(\Theta_1^T,\Theta_2^T,...,\Theta_n^T)^T$ and $X_0=(\Phi_1,\Phi_2,...,\Phi_M)^T$. Then $X_0$ is $M\times v$ dimensional. Let $Y=(\hat p_{11},\hat p_{12},...,\hat p_{1M},\hat p_{21},\hat p_{22},...,\hat p_{2M},...,\hat p_{nM})^T$, $X=I_n\otimes X_0$, $e=(e_{11},e_{12},...,e_{1M},e_{21},e_{22},...,e_{2M},...,e_{nM})^T$, $H=(1,1,...,1)_{1\times n}\otimes I_{v}$, $D_{v\times 1}=(\sqrt{d},0,...,0)^T$. Then the regression equations can be rewritten in a compact form:
$$\hat Y=X\Theta+e,$$
with a linear constraint
$$H\Theta=D,$$
which is the same form as (\ref{eqa06}) and (\ref{eqa07}), but with the dimensions of $\hat \Theta_{CLS}$ and $\hat \Theta_{LS}$ decreased from $nd^2$ to $nv$.

Before proceeding to the CLS solution, we introduce another amendment. In practical experiments, the types of the probe states are not always rich enough, and the resource number can be small. These limitations lead to large CLS error and thus unsatisfactory final errors. More specifically, physical estimations $\{\hat P_i\}$ always have the eigenvalues of $\hat P_i$ between 0 and 1, while bad nonphysical estimates usually make some of these eigenvalues far away from the region $[0,1]$, which indicates $||\hat \Theta_{CLS}||$ is too large. To avoid a CLS estimate that deviates seriously from the true value, we enforce a further requirement on the cost function of the linear regression process. Note that the original CLS problem is
\begin{equation}\label{eq49}
\min_{\hat \Theta}||\hat Y-X\hat\Theta||^2,\ \ s.t.\ \ H\hat\Theta=D.
\end{equation}
We now add an extra penalty item to modify (\ref{eq49}) as
\begin{equation}\label{eq50}
\min_{\hat \Theta}||\hat Y-X\hat\Theta||^2+\eta||\hat\Theta||^2,\ \ s.t.\ \ H\hat\Theta=D,
\end{equation}
where $\eta>0$. The new cost function is $\hat Y^T\hat Y-2\hat Y^TX\hat\Theta+\hat\Theta^T(X^TX+\eta I)\hat\Theta$. Hence, the new CLS solution is obtained by changing all the $X^TX$ items in (\ref{eqa08}) and (\ref{eqa09}) as $X^TX+\eta I$:
\begin{equation}\label{eq51}
\hat \Theta_{LS}=(X^TX+\eta I)^{-1}X^T\hat Y,
\end{equation}
and
\begin{equation}\label{eq52}
\begin{array}{rl}
\hat \Theta_{CLS}&=\hat \Theta_{LS}-(X^TX+\eta I)^{-1}H^T\\
&\ \ \ \ \cdot[H(X^TX+\eta I)^{-1}H^T]^{-1}(H\hat \Theta_{LS}-D).\\
\end{array}
\end{equation}
The modification from (\ref{eq49}) to (\ref{eq50}) is in essence \textit{Tikhonov regularization} \cite{convex book}, and the optimal parameter $\eta$ is usually difficult to determine by a fixed formula. Note that as the total resource number of all the probe states $N$ increases, $||\hat Y-X\hat\Theta||$ usually decreases, and $\eta$ should also decrease. We thus choose $\eta=10^3/N$ for simplicity. From the CLS solution (\ref{eq52}), we obtain the stage-1 estimate $\{\hat P_i\}$ which might not be positive semidefinite but satisfies all the other requirements.

The block diagonal structure of (\ref{eqa43}) implies that the detector is decoupled on the subspaces $\mathbb{C}_{d_1\times d_1}, \mathbb{C}_{d_2\times d_2},...,\mathbb{C}_{d_m\times d_m}$. We thus can perform the procedures of Sec. \ref{sec202}, \ref{sec203} and \ref{sec204} on these subspaces separately. Specifically, for each $1\leq j\leq m$, $\{\hat L^{(i)}_j\}_{i=1}^{n}$ is a set of Hermitian estimation on the space $\mathbb{C}_{d_j\times d_j}$ satisfying $\sum_{i=1}^n \hat L^{(i)}_j=I_{d_j}$. We thus employ \textit{difference decomposition}, \textit{stage-2 approximation} and \textit{unitary optimization} in Sec. \ref{sec202}- Sec. \ref{sec204} on $\{\hat L^{(i)}_j\}_{i=1}^{n}$ to obtain a set of physical estimation $\{\hat Q^{(i)}_j\}_{i=1}^{n}$ for each $j$. The final estimation is thus $\hat P_i=\hat Q^{(i)}_1\oplus \hat Q^{(i)}_2\oplus ...\oplus \hat Q^{(i)}_m$, which is physical and also satisfies the block-diagonal requirement.

\begin{remark}
An error upper bound similar to Theorem \ref{the1} can be given for this modified case. However, the upper bound requires that the form (\ref{eq49}) without the penalty item is employed and also that $N$ should be large enough. In practical experiments, $N$ is difficult to be arbitrarily large due to noise and imperfections. Hence, we do not present the similar error bound in this paper.
\end{remark}
\subsection{Experimental results}

We prepare two-mode coherent states for detector tomography by using an adequately attenuated continuous-wave (CW) fiber coupled laser as depicted in the yellow dashed box in Fig. \ref{fig}(a).
We express the general two-mode coherent state without global phase as $|\alpha,\beta \text{e}^{\text{i}\delta}\rangle$ ($\delta\in\mathbb{R}$, $\alpha,\beta\geq0$), which can be expanded in the photon number basis as
$$|\alpha,\beta \text{e}^{\text{i}\delta}\rangle=\exp[-\frac{1}{2}(\alpha^2+\beta^2)] \sum_{j,k}^{\infty}\frac{\alpha^j\beta^k\text{e}^{\text{i}k\delta}} {\sqrt{j!k!}}|j,k\rangle.$$
We can experimentally generate the above two-mode coherent states by attenuating the laser and rotating a QWP$_1$ and a half wave plate (HWP$_1$) after a PBS$_1$.
The probe states we used are the following 19 states:
\begin{equation}
\begin{array}{ccc}\hline
\alpha&\beta&\delta[\deg]\\
\hline
0.316&0.316&-135\\
0.316&0.316&-90\\
0.316&0.316&-45\\
0.316&0.316&0\\
0.316&0.316&45\\
0.316&0.316&90\\
0.316&0.316&135\\
0.316&0.316&180\\
0.447&0&-\\
0&0.447&-\\
0.194&0.112&-90\\
0.194&0.112&0\\
0.194&0.112&90\\
0.194&0.112&180\\
0.112&0.194&-90\\
0.112&0.194&0\\
0.112&0.194&90\\
0.112&0.194&180\\
0&0&-\\
\hline
\end{array}\nonumber
\end{equation}

We performed experiments for two different sets of detectors, denoted as Group I and Group II, respectively. We take $\eta=10^3/N$ for both groups. For the true value of Group I (experimental setting as Fig. \ref{fig}(b)), $P_1=L_1^{(1)}\oplus L_2^{(1)}\oplus L_3^{(1)}$, and we have $L_1^{(1)}=2.91\times 10^{-4}$,
\begin{equation}
L_2^{(1)}=\left(
\begin{array}{cc}
0.202&0.00109\text{i}\\
-0.00109\text{i}&0.202\\
\end{array}\right),\nonumber
\end{equation}
and
\begin{equation}
L_3^{(1)}=\left(
\begin{array}{ccc}
0.363&0.00123\text{i}&1.20\times10^{-6}\\
-0.00123\text{i}&0.363&0.00123\text{i}\\
1.20\times10^{-6}&-0.00123\text{i}&0.363\\
\end{array}\right).\nonumber
\end{equation}

For the true value of Group II (experimental setting as Fig. \ref{fig}(c)), we have $L_1^{(1)}=1.27\times 10^{-4}$,
\begin{equation}
L_2^{(1)}=\left(
\begin{array}{cc}
0.0763&-0.0440+0.0879\text{i}\\
-0.0440-0.0879\text{i}&0.127\\
\end{array}\right),\nonumber
\end{equation}
and
\begin{equation}
\begin{array}{l}
L_3^{(1)}=\ \ \ \ \ \ \ \ \ \ \ \ \ \ \ \ \ \ \ \ \ \ \ \ \ \ \ \ \ \ \ \ \ \ \ \ \ \ \ \ \ \ \ \ \ \ \ \ \ \ \ \ \ \ \ \ \ \ \ \ \ \ \ \ \ \ \ \ \\
\!\!\!\!\left(\!\!\!
\setlength{\arraycolsep}{1.6pt}
\begin{array}{ccc}
0.147&-0.0574+0.115\text{i}&0.00580+0.00773\text{i}\\
-0.0574-0.115\text{i}&0.184&-0.0543+0.109\text{i}\\
0.00580-0.00773\text{i}&-0.0543-0.109\text{i}&0.238\\
\end{array}\!\!\!\right)\!\!\!.\nonumber
\end{array}
\end{equation}
The error bars are at most 4\%, which are derived from the precisions of quantum efficiency measurements for each SNSPD.
\begin{figure}
\centering
\includegraphics[width=3.6in]{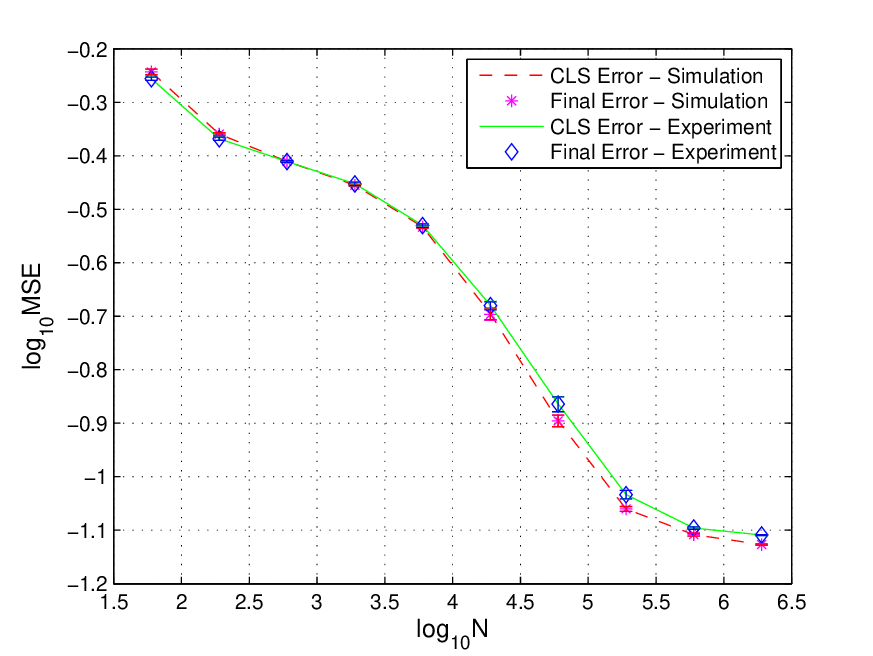}
\centering{\caption{Experimental and simulation results for Group I.}\label{exp1}}
\end{figure}

\begin{figure}
\centering
\includegraphics[width=3.6in]{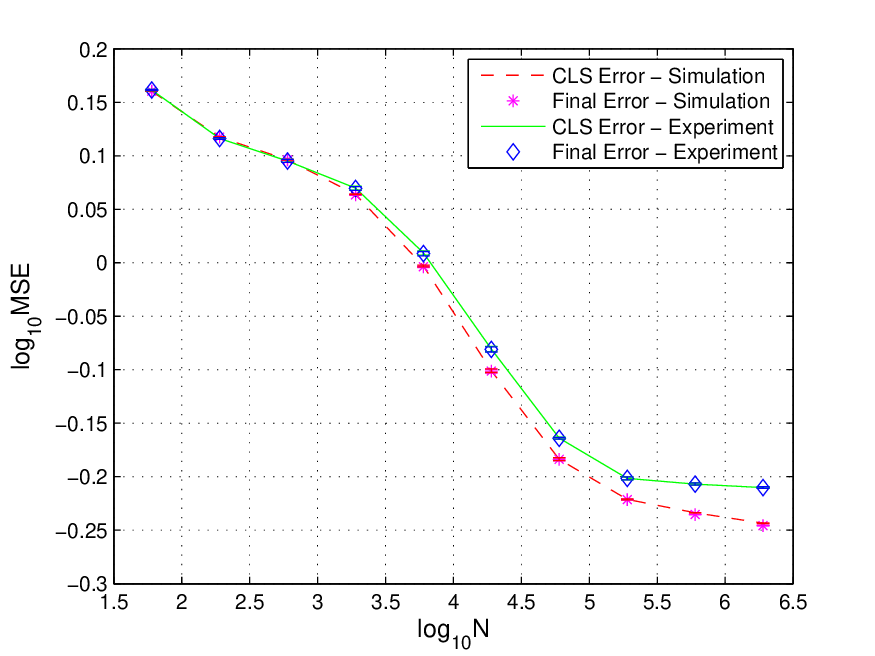}
\centering{\caption{Experimental and simulation results for Group II.}\label{exp2}}
\end{figure}

We record 100,000 measurement outcomes for each input state, and repeat it 6 times. By truncating the outcome records in the time axis we can obtain data for different resource numbers. We employ our modified algorithm to reconstruct the two sets of detectors, and show the results in Figs. \ref{exp1} and \ref{exp2}, respectively. We also plot the reconstruction results using simulated measurement data as a comparison. In Fig. \ref{exp1}, the simulation matches the experiment very well. The performance in Fig. \ref{exp2} is not as good as that for Group I, due to the influence of the nondiagonal elements with amplitudes significantly larger than zero.

\section{Conclusion}\label{secfinal}

In this paper, we have proposed a novel Two-stage Estimation (TSE) quantum detector tomography method. We analysed the computational complexity for our algorithm and established an upper bound for the estimation error. We discussed the optimization of the coherent probe states, and presented simulation results to illustrate the performance of our algorithm. Quantum optical experiments were performed and the results validated the effectiveness of our method.

\appendices

\section*{Acknowledgement}

The authors would like to thank Dr. Akira Sone for helpful discussions.

\ifCLASSOPTIONcaptionsoff
  \newpage
\fi

\end{document}